\documentclass[aps,prb,showpacs,twocolumn]{revtex4-1}

\usepackage[dvips]{graphicx}
\usepackage{amsmath,amssymb}
\usepackage{amsfonts}
\usepackage{color}

\usepackage[usenames,dvipsnames]{xcolor}

\begin{document}

\newcommand{\Eq}[1]{Eq.~\eqref{#1}}
\newcommand{\Eqs}[1]{Eqs.~\eqref{#1}}

\newcommand{\ui}{\text{i}}
\newcommand{\iexp}{\text{i}}
\newcommand{\im}{\text{Im}}
\newcommand{\re}{\text{Re}}

\newcommand{\uGHz}{\,\text{GHz}}
\newcommand{\uMHz}{\,\text{MHz}}
\newcommand{\uF}{\,\text{F}}
\newcommand{\uH}{\,\text{H}}
\newcommand{\um}{\text{m}}

\newcommand{\fq}{\left(\frac{\hbar}{2 e}\right)^2}
\newcommand{\fext}{f_{\text{ext}}}
\newcommand{\ELcav}{E_{L,\text{cav}}}
\newcommand{\df}{\delta\!f}

\newcommand{\dk}{\delta_k}
\newcommand{\enl}{\tilde \epsilon}
\newcommand{\znl}{\tilde \zeta}
\newcommand{\Dnl}{\tilde D}
\newcommand{\qnl}{\tilde q}
\newcommand{\etanl}{\tilde \eta}
\newcommand{\dth}{\sqrt{\epsilon^2-\Gamma^2}}
\newcommand{\ncvac}{n_c^{\text{vac}}}
\newcommand{\navac}{n_a^{\text{vac}}}

\title{Parametric resonance in tunable superconducting cavities}
\author{Waltraut Wustmann}
\author{Vitaly Shumeiko}
\affiliation{Chalmers University of Technology, S-41296 G\"oteborg, Sweden}
\date{27 February 2013}

\begin{abstract}
We develop a theory of parametric resonance in tunable superconducting cavities. The nonlinearity introduced by the SQUID attached to the cavity, and damping due to connection of the cavity to a transmission line are taken into consideration. We study in detail the nonlinear classical dynamics of the cavity field below and above the parametric threshold for the degenerate parametric resonance, 
featuring regimes of multistability and parametric radiation. We investigate the phase-sensitive amplification of external signals on resonance, as well as amplification of detuned signals, and relate the amplifier performance to that of linear parametric amplifiers. We also discuss applications of the device for dispersive qubit readout. 
Beyond the classical response of the cavity, we investigate small quantum fluctuations around the amplified classical signals. 
We evaluate the noise power spectrum both for the internal field in the cavity and the output field.
Other quantum statistical properties of the noise are addressed such as squeezing spectra, 
second order coherence, and two-mode entanglement.
\end{abstract}

\maketitle

\section{Introduction}
\label{sec:Introduction}

Parametric resonance is a fundamental physical phenomenon that is  encountered eventually in every area of science. In different disciplines, however, different facets of this rich phenomenon play a major role and are highlighted. Parametric instability and multistable regimes in nonlinear dynamics \cite{Nayfeh}, noise driven transitions among stable states in statistical physics \cite{DykMalSmeSil1998, Dykman2012}, wave mixing and frequency conversion in wave dynamics \cite{Sagdeev}  are topics of primary interest.
In electrical and optical engineering the low-noise properties of parametric amplifiers attract attentions, as well as  non-classical statistical properties of the electromagnetic field generated by parametric devices \cite{Clerk_review2010,Cav1982,SqueezingBook}. 

In superconducting electronics, the idea of using Josephson junctions  for quantum limited parametric amplification is under attention and development since the 1980s  \cite{YurDen1984,Yurke1988,YurETAL1989,YurETAL1990}. 
During the last years the field revived by challenges of quantum information technology. The circuit-QED design, initially proposed for qubit manipulation and measurement \cite{Blais2004,Schoelkopf2008}, was  employed for developing a variety of parametric devices \cite{
YamETAL2008,BergealNature2010,EichlerPRL2011,RocETAL2012,BergealPRL2012,FluETAL2012}. 

The circuit-QED approach is based on a combination of extended linear electromagnetic elements (transmission lines and resonators) 
with Josephson junctions as nonlinear lumped elements. The design is flexible, allowing for diverse methods of parametric pumping, phase preserving and phase sensitive amplification schemes, different numbers or input and output ports, distributed Josephson nonlinearities  \cite{CasLeh2007,CasETAL2008}.

The most of developed amplifiers are engineered in such a way that the dominant pump tone is 
sent through the same port as the signal, and parametric resonance is achieved by mixing them  
in nonlinear Josephson elements. A different method is available for tunable superconducting 
cavities~\cite{SanETAL2008,WalShu2006}. The device consists of a  resonator 
terminated with one (or more) dc-SQUID(s) that determines the reflection condition at the cavity 
edge and hence the cavity resonance spectrum. Parametric resonance is achieved by rapid 
modulation of a magnetic flux through the SQUID with an appropriate frequency. 
A number of interesting parametric effects have been observed with such a device: phase 
sensitive  amplification~\cite{YamETAL2008}, frequency conversion~\cite{AumentadoNaPh2011}, 
radiation and multistability regimes above the parametric threshold~\cite{WilETAL2010},    
quantum entanglement of output photons~\cite{MenzMarx2012}, generation of photons out of 
vacuum noise~\cite{WilETAL2011} - an analog of the dynamical Casimir effect~\cite{Moo1970,Nori2012}. 
 
In this paper we formulate a consistent theory of parametric resonance in a tunable 
superconducting cavity. We aim at a unified picture of the phenomenon below and above the 
parametric threshold. To this end we include into consideration the SQUID nonlinearity, and 
damping due to connection to a transmission line. The latter provides a stage for studying the 
parametric amplification.  We develop a full nonlinear description of the cavity resonance 
dynamics and the amplification effect in the classical limit, and study small quantum fluctuations 
of amplified and radiative fields. For certainty we consider parametric excitation of the main cavity mode $\omega_0$ by pumping with a frequency $\Omega$ close to twice the cavity resonance, $\Omega \approx 2 \omega_0$. 

The overall picture of nonlinear parametric resonance in the tunable cavity is rather rich 
and complicated. At very small pump strength the cavity intrinsic dynamics resembles the one of the Duffing oscillator \cite{Nayfeh} showing a bifurcation of the cavity response and bistability. However, the scattering of an external incidental wave is qualitatively different  from the Duffing case: the scattering is inelastic, the reflected wave undergoes amplification or deamplification  depending on the phase shift between the input tone and the pump (phase sensitive amplification). 

With increasing pump strength, the amplification effect increases, and at the same time the resonance narrows such that the bifurcation occurs at ever smaller input amplitudes. Eventually, while approaching the parametric threshold, the cavity response becomes nonlinear at any small input amplitude.

Further increase of the pump strength leads to an instability of the cavity zero-amplitude state and the formation of finite-amplitude states accompanied by stationary parametric radiation at the half frequency of the pump. The radiative states are bistable in a certain window of detuning of the pump frequency from the cavity resonance. Outside of this interval at red detuning the radiative states coexist with the stable zero-amplitude state (tristability), and the latter one becomes dominant at far red detuning. Remarkably all these multistable regimes have been observed in experiment with  a high quality tunable cavity \cite{WilETAL2010}. 

The multistability regimes are accompanied by random jumps among the stable states induced by thermal or quantum noise. These large amplitude fluctuations have small probability away from the bifurcation points and the parametric threshold, but become significant in the vicinity of these critical points (cf.~Ref.~\onlinecite{Dykman2012} and references therein). These effects are out of the scope of this paper, here we restrict to small quantum fluctuations around well defined classical states outside of the critical regions,
both below and above the parametric threshold. 

The bifurcation of the Duffing oscillator response is employed in Josephson bifurcation amplifiers (JBA) for dispersive qubit readout \cite{SidETAL2004, JBAreview2009}. This method also applies  to the parametric regime below the threshold (Josephson parametric bifurcation amplifier, JPBA). The novel feature here is the possibility to measure {\em amplitude} of the amplified probing tone, which exhibits strong dispersion with respect to the detuning near the threshold, and can be advantageous for high fidelity qubit readout.  

The parametric radiation above the threshold offers yet another strategy for the qubit readout based on the significant contrast between the strengths of the output radiation above the threshold and  the amplified noise below the threshold.

The paper is organized as follows. Sections \ref{sec:Device}, \ref{sec:FTCavity} and \ref{sec:Losses} are devoted to the development of the theoretical framework for describing parametric resonance in a high quality tunable cavity. In Sec.~\ref{sec:classical} we consider the nonlinear cavity response to a classical input signal below and above the parametric threshold, and in Sec.~\ref{sec:application} apply the results for the analysis of parametric amplification and  methods of dispersive qubit readout. Section  
\ref{sec:quantum} is devoted to the analysis of quantum fluctuations.

\section{Circuit Lagrangian}
\label{sec:Device}

\begin{figure}[tbh]
\centering
\includegraphics[width=0.9\columnwidth]{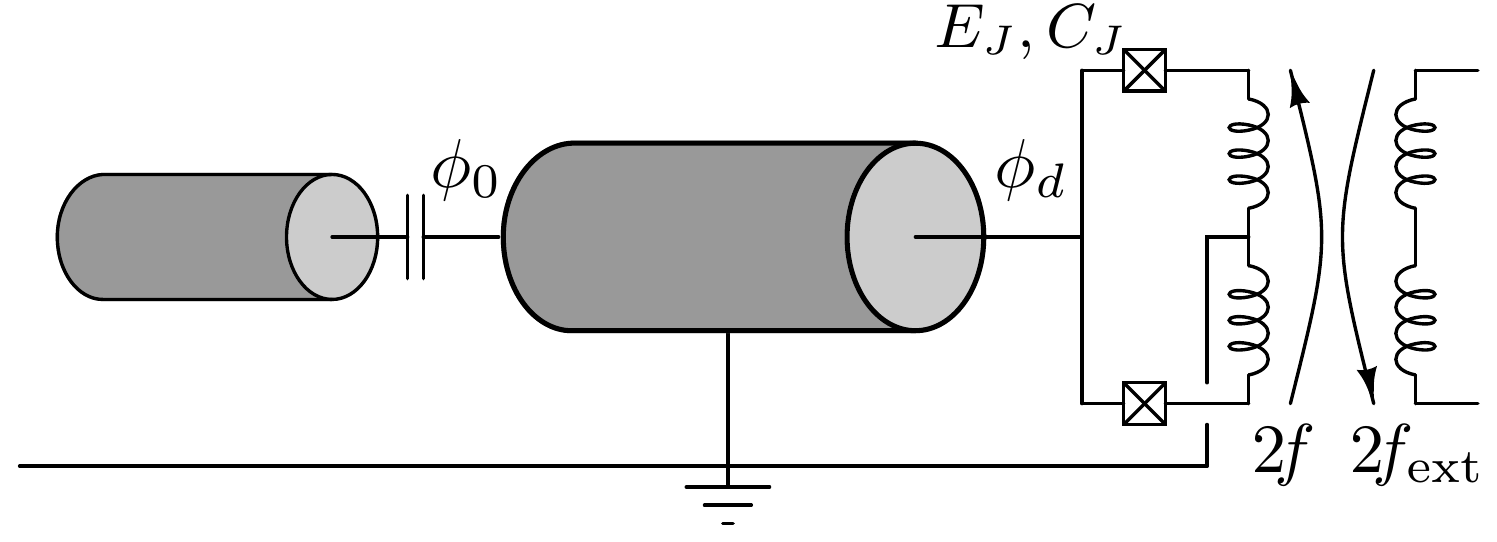}
\caption{
Sketch of a tunable cavity device: the $\lambda/4$-cavity is terminated by a dc SQUID at the right end,
and is capacitively coupled to a transmission line at the left end;
the SQUID is flux biased (phase $2\!f$)  via inductive coupling to a flux line imposing a driving phase $2\!f_{\text{ext}}$; $\phi_d = \phi(d,t)$ and $\phi_0 = \phi(+0,t)$ are the phase values at the right and left ends of the cavity, respectively. An incidental signal fed in from the transmission line is reflected, separated from the input, and then analyzed.
}
\label{fig:device}
\end{figure}

The device we study is sketched in Fig.~\ref{fig:device}. Its main part is a tunable 
superconducting strip line cavity terminated with a SQUID~\cite{SanETAL2008,WalShu2006}. The cavity is weakly coupled 
to a transmission line that feeds an external microwave signal in and provides means for probing 
the field inside the cavity. The cavity is a spatially extended system of length $d$ with inductance $L_0$ and  capacitance $C_0$ per unit length,
and the cavity state is characterized by the superconducting phase field $\phi(x,t)$.
We use the Lagrangian formalism~\cite{YurDen1984,Devoret2004,WalShu2006} 
to describe the nonstationary dynamics of $\phi(x,t)$. 

The Lagrangian of the entire device consists of the 
sum of the Lagrangians of the cavity, transmission line, and the coupling, 
\begin{equation}
\mathcal{L}[\phi] = \mathcal{L}_{\text{cav}} + \mathcal{L}_{TL} + \mathcal{L}_{c}
\,.
\end{equation}
The Lagrangian of the cavity in its turn consists of the Lagrangian of the bare cavity $\mathcal{L}_{\text{cav}}^{(0)}$,
and the Lagrangian of the SQUID  $\mathcal{L}_S[\phi_d]$,
\begin{eqnarray}
 \mathcal{L}_{\text{cav}} 
&=& \mathcal{L}_{\text{cav}}^{(0)}[\phi] + \mathcal{L}_{S}[\phi_d] \nonumber \\
\label{eq:Lcav_x_cos_orig0}
&=&  \fq \frac{C_0}{2} \int_0^d d x \left(\dot \phi^2 - v^2 \phi'^2 \right) \\
&+& \left[ \fq \frac{2 C_J}{2} \dot \phi_d^2 
 + 2 E_J \cos{f(t)} \cos\phi_d
 \right]
\nonumber \,.
\end{eqnarray}
Here $v = 1/\sqrt{L_0 C_0}$ is the field propagation velocity, $\phi_d(t) = \phi(d,t)$ is the boundary value of the cavity field at the SQUID, and $f(t)$ is the phase across the SQUID controlled by external magnetic flux, see Fig.~\ref{fig:device}. 
The SQUID is assumed symmetric for simplicity, with two identical Josephson junctions, 
each having a Josephson energy $E_J$ and a capacitance $C_J$. The phase $f(t)$ 
appears in \Eq{eq:Lcav_x_cos_orig0} as an external time-dependent parameter that is able to 
excite parametric resonance. 
In fact it is a dynamical variable that describes, together with the variable $\phi_d$, the dynamics of two coupled Josephson oscillators of the SQUID driven by the external electromagnetic field $\fext(t)$. 
In Appendix \ref{sec:SQUID} we show that in the limit of small $\phi_d \ll 1$
the $f$-oscillator decouples from the $\phi_d$-oscillator. 
Moreover, for experimentally relevant circuit parameters,  the $f$-oscillator follows the drive field adiabatically because the resonance frequency of the $f$-oscillator is large compared to a typical resonance frequency of the cavity. 
A detailed derivation of \Eq{eq:Lcav_x_cos_orig0} and the connection of $f(t)$ to the 
external field $\fext(t)$ is provided in Appendix~\ref{sec:SQUID}. 

We assume here that the controlling field $f(t)$ is composed of a constant biasing part $F$ and a small harmonic oscillation with amplitude $\df \ll 1$,
\begin{equation}
f(t) = F + \df \cos\Omega t
\,.
\end{equation}

It is worth mentioning that the constraint 
$\phi_d \ll 1$ is essential, otherwise the two Josephson oscillators become coupled and exhibit complex, even chaotic behavior under external drive~\cite{LicLie1992}. 

Proceeding to the other components of the device, we suppose the transmission line to have the same characteristic parameters $C_0$ and $L_0$ as the cavity,
\begin{equation}\label{eq:LTL}
 \mathcal{L}_{TL}[\phi_{TL}] = \fq \frac{C_0}{2} \int_{-\infty}^0 d x \left(\dot \phi_{TL}^2 - v^2 \phi_{TL}'^2 \right)
\,.
\end{equation}
The capacitive coupling is described with the Lagrangian
\begin{eqnarray}\label{eq:LcavTL}
\mathcal{L}_{c}  
 = \fq \frac{C_c}{2} \left(\dot \phi_0 - \dot \phi_{TL,0} \right)^2 
\,,
\end{eqnarray}
where  $\phi_0 = \phi(+0,t)$ and $\phi_{TL,0} = \phi_{TL}(-0,t)$ 
are the field values at the different sides of the coupling capacitor $C_c$.

\section{Parametric dynamics of closed cavity}
\label{sec:FTCavity}

We first consider the cavity decoupled from the input line, $C_c = 0$. The goal will be to identify the cavity frequency spectrum and investigate the parametric resonance.

\subsection{Cavity modes}
\label{subsec:CavModes}

The Lagrangian $\mathcal{L}_{\text{cav}}$, Eq.~\eqref{eq:Lcav_x_cos_orig0}, 
explicitly contains two dynamical variables, the  phase field $\phi(x,t)$,
and its boundary value  $\phi_d(t)$.
Variation of the associated action with respect to $\phi(x,t)$ leads to the wave equation,
\begin{equation}
\ddot \phi - v^2 \phi'' = 0
\,,
\end{equation}
supplemented by the boundary condition $\phi'_0=0$ at the open end of the cavity.
Variation with respect to the boundary value $\phi_d(t)$ yields the boundary condition,
\begin{eqnarray}\label{eq:EOM_phid}
\frac{\hbar^2}{E_C} \ddot \phi_d + 2 E_J \cos{f(t)} \sin\phi_d + \ELcav d \phi'_d  &=& 0
\,,
\end{eqnarray}
where $E_C = (2e)^2/(2 C_J)$ and $\ELcav = (\hbar/2e)^2 (1/L_0 d)$.

Under static biasing, $\df=0$, the linearized boundary condition of \Eq{eq:EOM_phid}
determines the set of cavity eigen modes~\cite{WalShu2006}, 
\begin{eqnarray}
\label{eq:phi_n}
\phi_n(x,t) \propto e^{\pm \iexp \omega_n t} \cos{k_n x}\,, \quad \omega_n = v k_n \\
\label{eq:dispersion}
(k_n d)\tan{k_n d}  =   \frac{2 E_J \cos{F} }{\ELcav}  - \frac{2 C_J }{C_0 d}(k_n d)^2
\,.
\end{eqnarray}
The frequency spectrum $\omega_n$ is non-equidistant, and can be tuned by varying the bias $F$.

Although the first term at the rhs of \Eq{eq:dispersion} can in principle be tuned to zero, at $F=\pi/2$, in practice  it  dominates over the second term, at least for the lowest cavity modes, by virtue of the large parameter $\omega_J / \omega_n \gg 1$, where $\omega_J = \sqrt{2 E_J E_C}/\hbar$ is the Josephson plasma frequency. 
Indeed, given typical experimental values, 
$E_J/\hbar \approx 4500 \uGHz$ and $E_C/\hbar \sim 10 \uGHz$, the plasma frequency is $\omega_J \approx 300 \uGHz$, while the cavity fundamental frequency is $\omega_0 \sim 40 \uGHz$, i.e., by one order of magnitude smaller (for typical cavity parameters\cite{SanETAL2008,WilETAL2010} $L_0 \sim 4 \cdot 10^{-7} \uH/\um$, $C_0 \sim 2\cdot 10^{-10} \uF/\um$, and $d \approx \lambda/4$). 

\begin{figure}[tb]
\centering
\includegraphics[width=\columnwidth]{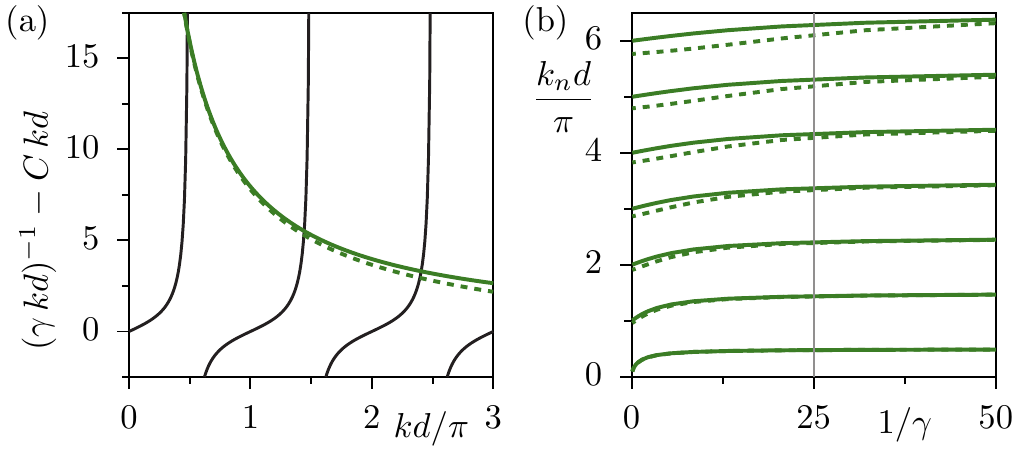}
\caption{
Cavity spectrum:
(a) graphical solution of \Eq{eq:dispersion} for $1/\gamma = 25$, and $C_J=0$ (solid) 
and $C = 2C_J/(C_0 d) = 0.05$ (dashed);
(b) cavity spectrum $k_n d$ vs.~$1/\gamma$ according to \Eq{eq:dispersion}, 
the vertical line indicates the value $1/\gamma = 25$ used in (a).
} 
\label{fig:spectrum}
\end{figure}

Furthermore, the cavity inductive energy is typically small, $\ELcav/\hbar \sim 400 \uGHz$, compared to the Josephson energy $2 E_J$. Taking advantage of this relation, and neglecting the capacitive term in \Eq{eq:dispersion}, we get the approximate solutions
\begin{eqnarray}\label{eq:w0_approx}
k_0 d &\approx& \frac{\pi}{2} \left(1 -\gamma \right) \approx{\pi\over 2}\,, \quad 
\gamma =\frac{\ELcav}{2 E_J \cos{F} }\ll 1\,,
 \\
k_n &\approx& k_0 +\pi n/d
\,.
\end{eqnarray}
The solutions of the spectral equation \eqref{eq:dispersion} are graphically illustrated 
in Fig.~\ref{fig:spectrum}(a), while Fig.~\ref{fig:spectrum}(b) shows the cavity spectrum as a function of the parameter $1/\gamma$.

\subsection{Cavity Hamiltonian}
\label{subsec:Hamiltonian}

The Lagrangian formalism is sufficient for analyzing the classical parametric resonance. To describe the quantum dynamics the Hamiltonian approach is more convenient. We derive the cavity Hamiltonian by expanding the cavity field over the complete set of cavity eigen modes,
\begin{equation}\label{eq:modeexpansion}
\phi(x,t) = {2e \over \hbar} \sqrt{2 \over C_0 d} \sum_n q_n(t) \cos k_nx
\,,
\end{equation}
where $q_n(t)$ are  time-dependent coefficients, and $k_n$ obey \Eq{eq:dispersion}. 
Using expansion~\eqref{eq:modeexpansion}  and noticing that the set of functions $\cos k_nx$ is non-orthogonal, we present the Lagrangian (\ref{eq:Lcav_x_cos_orig0}) after some algebra  in the form,
\begin{eqnarray}\label{eq:Lcav_n_nlin}
 \mathcal{L}_{\text{cav}}
= \frac{1}{2} \sum_n \left[ M_n \dot q_n^2 - M_n \omega_n^2 q_n^2 \right] - V(q_n,t)
\,.
\end{eqnarray}
Here the ``masses'' of the mode oscillators are given by the expressions,
\begin{equation}\label{eq:Mn}
M_n = 1 + {\sin{2k_n d} \over 2k_nd} + {4C_J\over dC_0}\cos^2k_nd
\,, \quad M_0\approx 1
\,,
\end{equation}
and 
\begin{equation}
V(q_n,t) = 
- 2 E_J \left[\cos f(t)  \cos\phi_d + \cos F \phi_d^2/2 \right]
\end{equation}
is a nonstationary nonlinear potential that mixes the eigen modes (see Appendix \ref{sec:Lcav_n} for  details of the derivation). 

It is convenient to absorb the factors $M_n$ and $\omega_n$ into the rescaled coordinate,
\begin{equation}
\sqrt {M_n\omega_n}\,q_n\; \rightarrow \; q_n
\,,
\end{equation}
and redefine the mode expansion in \Eq{eq:modeexpansion} accordingly.
Then introducing the conjugated momenta,
$p_n = \partial \mathcal{L}\big/\partial \dot q_n =  \dot q_n/\omega_n$, 
we arrive at the cavity Hamiltonian,
\begin{eqnarray}\label{eq:Hcav}
 H_{\text{cav}}(q_n, p_n) &=& {1\over 2}\sum_n \omega_n \left( p_n^2 + q_n^2 \right) 
 + V(q_n,t)
\,.
\end{eqnarray}
%

\subsection{Resonance approximation}
\label{sec:parametric}

For small pumping amplitudes and weak non-linearity, the potential $V(q_n,t)$  in \Eq{eq:Hcav} could be considered perturbatively. 
However, the perturbative approach does not apply to the case of parametric resonance, when 
the pumping frequency matches an algebraic sum of the cavity eigen frequencies, $\Omega 
\approx \omega_n \pm \omega_m$. In this case the corresponding cavity modes are strongly 
mixed and undergo complex time evolution. 
A particular case is the degenerate parametric resonance for $m=n$. In this paper we consider 
for certainty the degenerate parametric resonance of the fundamental mode, $\Omega \approx  
2\omega_0$. The method outlined below is straightforwardly extended to a non-degenerate parametric resonance.  

First we perform a canonical transformation corresponding to a transition to the rotating frame with frequency $\Omega/2$. This is conveniently done in terms of a complex variable,
\begin{equation}\label{eq:def_a}
a_n = (q_n + \ui p_n)/\sqrt{ 2 \hbar } 
\,,
\end{equation}
for which the transformation reads $a_n(t) = e^{-i\Omega t/2} A_n(t)$. 
The equations of motion for the amplitudes $A_n(t)$ read,
\begin{equation}
\dot A_n = -\ui(\omega_n-\Omega/2)A_n 
- {\ui \over \sqrt{2\hbar}} {\partial V(q_n,t)\over \partial q_n} e^{\iexp\Omega t/2}
\,.
\end{equation}
At this point we take advantage of small values of the pumping amplitude, $\df \ll 1$, and the field amplitude, $\phi_d \ll 1$, and  expand the potential $V(q_n,t)$ in powers of these small parameters, keeping only the first non-vanishing terms, 
\begin{equation}\label{eq:approximateV}
V(q_n,t)\approx 
 -\left(E_J \df \sin F \cos\Omega t \right) \phi_d^2 
 - {E_J\over 12} \cos F \, \phi_d^4 
\,.
\end{equation}
Close to the resonance, $\Omega/2 - \omega_0 = \delta \ll \omega_n, \;\omega_n-\omega_m$,
the variable $A_0$ depends slowly on time while all the other variables contain rapid time oscillations. After averaging over these oscillations we arrive at the shortened equation of motion for $A_0$ (we skip the mode index $0$ below),
\begin{equation}\label{eq:EOM_A_cl_closed}
\dot A - \ui \delta A - \ui \epsilon A^\ast - \ui \alpha |A|^2 A = 0
\,,
\end{equation}
with the parameters
\begin{eqnarray}
\label{eq:def_epsn}
 \epsilon &=& 
 \frac{\df \omega_0 \tan F}{2\gamma} \frac{\cos^2k_0d}{M_0 (k_0d)^2} 
\,, \\
\label{eq:def_alphan}
 \alpha &=& \frac{\hbar \omega_0^2}{2\gamma\,\ELcav } \left(\frac{\cos^2k_0d}{M_0 (k_0d)^2}\right)^2
\,.
\end{eqnarray}

When applying the canonical transformation to coordinate and momentum,  
$A = (Q+\ui P)/\sqrt{2\hbar}$,
and averaging over fast oscillations, the cavity Hamiltonian is cast into the form,
\begin{eqnarray}\label{eq:Hcav_RF_QP}
 \!\!\!\! H_{\text{cav}}(Q, P) = \frac{\epsilon-\delta}{2} P^2 \! - \! \frac{\epsilon+\delta}{2} Q^2
\!\! - \! \frac{\alpha}{8 \hbar}\! \left(Q^2 + P^2\right)^2\!
.
\end{eqnarray}
This Hamiltonian corresponds to the metapotential of the parametric lumped element oscillator~\cite{DykMalSmeSil1998}, i.e.~the degenerate parametric resonance in the cavity is mapped on the one in a lumped element oscillator. The mapping is defined by \Eqs{eq:def_epsn} and 
(\ref{eq:def_alphan}), where the effective pump strength $\epsilon$, and the nonlinearity coefficient $\alpha$ are expressed through  generic cavity parameters. 

According to the experimental  values discussed in Sec.~\ref{subsec:CavModes},
parameter $\gamma$ in \Eq{eq:w0_approx} is estimated as $\gamma \sim 4 \cdot 10^{-2}$.
For such a small value of $\gamma$, the parameters $\epsilon$ and $\alpha$ are approximated, using the spectral equation (\ref{eq:dispersion}),
\begin{eqnarray}
\epsilon/\omega_0 &\approx& \gamma \df \tan{F} /2  \ll \df \,, \\
%
\alpha/\omega_0 &\approx& \gamma^{3} \frac{\hbar \omega_0}{2 \ELcav} 
= \gamma^{3}  { \pi^2 Z_0 \over 2 R_q} \lesssim 10^{-5}
\,, 
\end{eqnarray}
where $Z_0 = \sqrt{L_0/C_0}$ is the cavity impedance and $R_q = h / 2e^2$ is the quantum resistance. 

It follows from these estimates that the effective pump strength $\epsilon$ is substantially reduced compared to the amplitude of the phase modulation in the SQUID, and the effective nonlinearity of the cavity oscillator is significantly smaller than the underlying bare nonlinearity of the SQUID oscillator ($\alpha = \omega_J/6$ for the Josephson potential). 
These remarkable properties result from the fact that the cavity is almost shortcut to the ground at the edge $x=d$ by virtue of large Josephson energy in \Eq{eq:EOM_phid}
($\gamma \ll 1$),
hence the boundary value of the field amplitude $\phi_d$ is small.  

The small values of the effective oscillator parameters are essential for the validity of the resonance approximation. The latter requires  the evolution of $A$ to take place on a time scale
much larger than the period of the cavity fundamental mode, $1/\omega_0$, over which the initial Hamiltonian is averaged.

It is instructive to express the  constraints earlier imposed on the phases, $\phi_d, \df \ll 1$,
 in terms of the amplitude $A$ and the pump strength $\epsilon$,
\begin{equation}\label{eq:constraint_A}
|A| \ll  \sqrt {R_q / Z_0  \gamma^2}, \quad \epsilon \ll \gamma\omega_0
\,,
\end{equation}
or equivalently, $\alpha|A|^2, \, \epsilon \ll \gamma\omega_0 \ll \omega_0$.  In other words,  the constraints (\ref{eq:constraint_A}) are more stringent than the ones required for the resonance approximation, $\alpha|A|^2,\, \epsilon \ll \omega_0$. On the other hand, these constraints provide sufficient room for the pumping strength to be increased above the parametric threshold beyond the resonance width $\Gamma$  (see \Eq{eq:def_Gamma} in the next section), 
$\alpha|A|^2, \, \epsilon \sim \Gamma \ll \gamma\omega_0$, for a high quality cavity.

In most of our calculations we  restrict to the lowest order $\df$-dependence in \Eq{eq:approximateV}, however, in some cases it is useful to keep higher order terms. In particular, the second order term $\propto \df^2$  will introduce, after averaging over time, a nonlinear shift of the resonator frequency, proportional to $\epsilon^2$. This shift is evaluated in  \Eq{eq:pump_frshift} in Appendix~\ref{sec:SQUID}, and in terms of the effective pump strength it reads,
\begin{equation}\label{eq:omega_shift}
{\omega_0\,(\epsilon) - \omega_0 \over \omega_0} 
\approx
-{\epsilon^2 \over \gamma\omega_0^2\tan^2 F}  
\,.
\end{equation}
This shift could be used in practice for evaluating the actual magnitude of the pump power acting upon the SQUID, which is usually not known. Also, it causes quenching of the parametric  instability
at large pump strength, as will be shown in Sec.~\ref{sec:paramresonance}.

\section{Cavity coupled to transmission line}
\label{sec:Losses}

The parametric effect in the closed cavity is an idealization. The connection to the external transmission line gives rise to the qualitatively important new features: firstly, the cavity field is allowed to leak out of the cavity, giving rise to the cavity damping, and secondly, an external electromagnetic signal can be fed into the cavity and amplified.  Our aim in this section will be to include these features into \Eq{eq:EOM_A_cl_closed}, and derive the relation between the input and output fields, thus preparing the framework for the further investigation of parametric amplification. Our derivation closely follows the input-output theory \cite{YurDen1984, ColGar1984}, (see also illuminative derivations in Refs.~\onlinecite{Clerk_review2010,Yurke_DruFic2004}). 

Aiming at the analysis of the quantum dynamics of the open cavity, we describe the field in the transmission line in terms of spatial modes, similar to \Eq{eq:modeexpansion} for the cavity,
\begin{equation}\label{eq:phiTL}
\phi_{TL}(x,t) = {2e\over\hbar}\sqrt{ 2\over C_0 \pi}
\int_0^\infty {d k \over \sqrt{\omega_k}} q_k(t) \cos kx
\,, 
\end{equation}
with $\omega_k = vk$.
Opening of the cavity invokes also an additional set of modes $\propto \sin{k x}$, however, in the weak coupling limit these modes do not contribute in the main approximation and are neglected here. 

Focusing on the effect of cavity damping at weak coupling, we will  only keep the cross term in the coupling Lagrangian (\ref{eq:LcavTL}), 
\begin{equation}\label{eq:LcavTL_bare}
\mathcal{L}_{c} = - \fq C_c \dot \phi_0 \dot \phi_{TL,0}
\,,
\end{equation}
and neglect the quadratic terms, thus neglecting small corrections to the kinetic energies.
With this simplification, and retaining only the fundamental mode field in the cavity Lagrangian, we write the total Lagrangian in the form,
\begin{eqnarray}\label{eq:coupl_L}
\mathcal{L} &=& 
{1\over2}\left({\dot q^2\over \omega_0} - \omega_0 q^2 \right) - V(q,t) + {1\over2} 
\int_0^\infty d k \left({\dot q_k^2\over \omega_k} - \omega_k q_k^2 \right) 
\nonumber\\
&&-{C_c\over C_0\sqrt {M_0\pi d} } \int_0^\infty {d k\over \sqrt{\omega_0\omega_k}}\,
\dot q \dot q_k
\,.
\end{eqnarray}
The corresponding Hamiltonian reads, to first order of the weak coupling ($C_c \ll C_0 d$),
\begin{eqnarray}\label{eq:coupl_H}
\mathcal{H} &=&
{\omega_0\over2} \left( p^2 + q^2 \right) \! +\! V(q,t) 
+ {1\over2}\int_0^\infty \!\!\! d k \,\omega_k \left( p_k^2 + q_k^2 \right) 
\nonumber \\
&& + {C_c\over C_0\sqrt {M_0 \pi d}} \int_0^\infty d k \sqrt{\omega_0\omega_k} \,p p_k
\,.
\end{eqnarray}

Repeating the derivation of the previous section we derive coupled equations of motion for the cavity amplitude $a$ and the spectral amplitudes 
of the transmission line, $a_k = (q_k+\ui p_k)/\sqrt{2\hbar}$,
\begin{eqnarray}
\!\!\!\!&\ui & \dot a = \omega_0 a + {\partial V(q_n,t)\over \sqrt{2\hbar} \,\partial q_n} 
+ \ui \sqrt{2\Gamma_0 \over\pi\hbar k_0}   \int_0^\infty d k \sqrt{\omega_k} \, p_k \\
\label{eq:EOM_ak}
\!\!\!\!&\ui &\dot a_k = \omega_k a_k  +  \ui \sqrt{2\Gamma_0 \over\pi\hbar k_0}  
\sqrt{\omega_k} \, p 
\,.
\end{eqnarray}
Here we introduced the cavity damping rate, 
\begin{equation}\label{eq:def_Gamma}
 \Gamma_0 =\omega_0 \left({C_c \over C_0 d }\right)^2 { k_0d \over M_0} 
\,.
\end{equation}

Near the parametric resonance the equations of motion for the slow variables, 
$A(t) = e^{\iexp \Omega t/2} a(t)$, and $A_k(t) =  e^{\iexp \Omega t/2} a_k(t)$, take the form, 
after averaging over rapid time oscillations, 
\begin{eqnarray}\label{eq:A1}
\!\!\!\!  &\ui& \dot A + (\delta + \alpha |A|^2)A + \epsilon A^\ast
=  \sqrt{\Gamma_0 \over\pi k_0}  \int_{0}^\infty \!\!\! d k \,\sqrt{\omega_k} A_k \\
\label{eq:Ak}
\!\!\!\! &\ui &\dot A_k - \delta_k A_k  =  \sqrt{\Gamma_0 \omega_k\over\pi k_0}  A
\,, 
\end{eqnarray}
with $\delta_k = \omega_k -\Omega/2$.

We eliminate the transmission line modes from \Eq{eq:A1}, invoking the solutions of \Eq{eq:Ak},
\begin{eqnarray}
\label{eq:Ak_solt0}
A_k(t) 
&=& A_k(t_0) e^{-\iexp \delta_k (t-t_0)}  \\
&-& \ui \sqrt{\Gamma_0 \omega_k\over\pi k_0} 
\int_{t_0}^t d t'  e^{-\iexp \delta_k (t-t')} A(t'), \quad t_0<t 
\,,
\nonumber
\end{eqnarray}
with initial conditions $A_k(t_0)$ at time $t_0 < t$, and substituting it into \Eq{eq:A1}.
Within the resonance approximation, the factor $\sqrt{\omega_k}$ in the integrand is to be replaced 
with $\sqrt{\omega_0}$, and the integration over the wave vector $k$ to be extended to the entire axis. 
After making these approximations we arrive at the Langevin equation for the cavity amplitude,
\begin{equation}\label{eq:EOM_A_cl}
\ui \dot A  + \delta A + \epsilon A^\ast + \alpha |A|^2 A  
+\ui \Gamma_0 A = \sqrt{2\Gamma_0} B(t) 
\,,
\end{equation}
with the input flux amplitude
\begin{eqnarray}\label{eq:Bt}
B(t) &=& {1 \over \sqrt{2\pi v}} \int_{-\infty}^\infty 
d \delta_k   \, A_k(t_0) e^{-\iexp \delta_k (t-t_0)} 
\,.
\end{eqnarray}
The amplitude $B(t)$ is associated, as shown in Appendix \ref{sec:Langevin}, with the incident (right-going) wave in the transmission line, $B(t-x/v)$, taken at the boundary $x=0$.

The solution of \Eq{eq:Ak} can be equivalently expressed in terms of the amplitude at a future time, $A_k(t_1)$, $t_1>t$, which defines the output flux amplitude $C(t)$ via a relation similar to \Eq{eq:Bt} with $t_1$ substituting for $t_0$. This output amplitude is associated with the reflected (left-going) wave in the transmission line, $C(t+x/v)$, taken at  $x=0$ (Appendix \ref{sec:Langevin}). The relation between the output and input amplitudes reads, 
\begin{equation}\label{eq:relation_ABC}
C(t) = B(t) - \ui \sqrt{2\Gamma_0} A(t)
\,.
\end{equation}

The parametric pumping couples the cavity field amplitude $A$ and its complex conjugate, and it is convenient to rewrite \Eq{eq:EOM_A_cl} in the matrix form,
\begin{eqnarray}\label{eq:EOM_a_matrix}
\frac{d}{d t} 
\left( \begin{array}{l} \phantom{-} \ui A \\ - \ui A^{*} \end{array} \right)
+ \mathcal{A} 
\left( \begin{array}{l} A \\ A^{*}  \end{array} \right) 
=  \sqrt{2\Gamma_0}
\left( \begin{array}{l}  B \\ B^{*}   \end{array} \right)
\,,
\end{eqnarray}
where
\begin{eqnarray}\label{eq:matA}
\mathcal{A} &=& 
\left(\begin{array}{cc}
\zeta+\ui\Gamma  & \epsilon \\
\epsilon & \zeta -\ui\Gamma 
\end{array}\right) , \nonumber\\ 
\zeta &=& \delta + \alpha |A|^2
\,.
\end{eqnarray}

The conservative part of the dynamics in \Eqs{eq:EOM_A_cl}, \eqref{eq:EOM_a_matrix} is determined by the effective Hamiltonian
\begin{equation}\label{eq:Hcav_RF_QP_B}
 H(Q, P) = H_{\text{cav}} + 2 \sqrt{\Gamma_0} |B| (\cos\theta_B  Q + \sin\theta_B P)
\,,
\end{equation}
with $H_{\text{cav}}(Q, P)$ from \Eq{eq:Hcav_RF_QP}, 
and $\theta_B$ being the phase shift between the input amplitude $B=|B|e^{\iexp \theta_B}$ and the pump.

Besides the damping $\Gamma_0$  associated with the opening of the cavity, there might also 
be internal losses in the cavity, e.g.~caused by the cavity resistance. 
A way to account for these losses is a model with a fictitious transmission line coupled to the cavity, that acts as a scattering channel with a noisy input amplitude $B_R(t)$ and an associated damping rate 
$\Gamma_R$. This would lead to an enhanced damping rate, $\Gamma = \Gamma_0 + \Gamma_R$, at the lhs of \Eq{eq:EOM_A_cl}, and also introduce an additional input term, $\sqrt{2\Gamma_R} B_R(t)$, at the rhs of this equation.

The damping effect results in the broadening of the resonance, and if the resonance becomes sufficiently broad, higher cavity modes might also be excited, despite the non-equidistant property of the cavity spectrum. In this case,  the isolated mode dynamics of \Eq{eq:EOM_A_cl} would be replaced by a more complex dynamics of  parametrically excited coupled  modes.
To ensure the validity of the single-mode approximation, the 
condition $|\omega_0 \pm \Gamma/2 + \Omega - \omega_1| > \Gamma_1/2$ must be met,
where $\Gamma_1$ is the resonance width of the first cavity mode. For $\Omega = 2\omega_0$, and the cavity spectrum given by \Eq{eq:dispersion}
and parameters of Sec.~\ref{subsec:CavModes}, the anharmonicity is of the order,  
$\omega_0+\Omega/2-\omega_1 = 3\omega_0-\omega_1 \approx 10^{-3} \omega_0$. This implies that the cavity quality factor $Q=\omega_0/\Gamma $ should  not be less than $Q\gtrsim 10^3 $. This corresponds to a small coupling capacitance in \Eq{eq:def_Gamma}, $C_c/C_0d \sim 10^{-2}$, assuming that the internal losses are not dominant, $\Gamma_R < \Gamma_0$.

\section{Classical cavity response and radiation}
\label{sec:classical}

In this section we analyze the cavity response to a noiseless classical input signal. 
We consider harmonic inputs, which 
have the form $B(t)= Be^{-\iexp \Delta t}$, where $\Delta = \omega-\Omega/2$ is the detuning of the input signal from the half frequency of the pump. For the input frequency  $\Delta=0$ the cavity response is stationary, and it can be fully analyzed in the nonlinear regime. 
For detuned inputs, we restrict to small input amplitudes; at large amplitudes the nonlinear response becomes complex and exhibits a transition to a chaotic regime.

\subsection{Parametric resonance in absence of input signal}
\label{sec:paramresonance}

We start with the analysis of the intrinsic parametric resonance in the cavity in the absence of input signals, $B(t)=0$.
Due to the damping, any initial cavity state
evolves towards one of the steady states that define the picture of the parametric resonance. 
These steady states depend crucially on the pump strength $\epsilon$, and also on the detuning of the pump frequency from the cavity resonance, $\delta = \Omega/2-\omega_0$.

\begin{figure}[tb]
\centering
\includegraphics[width=\columnwidth]{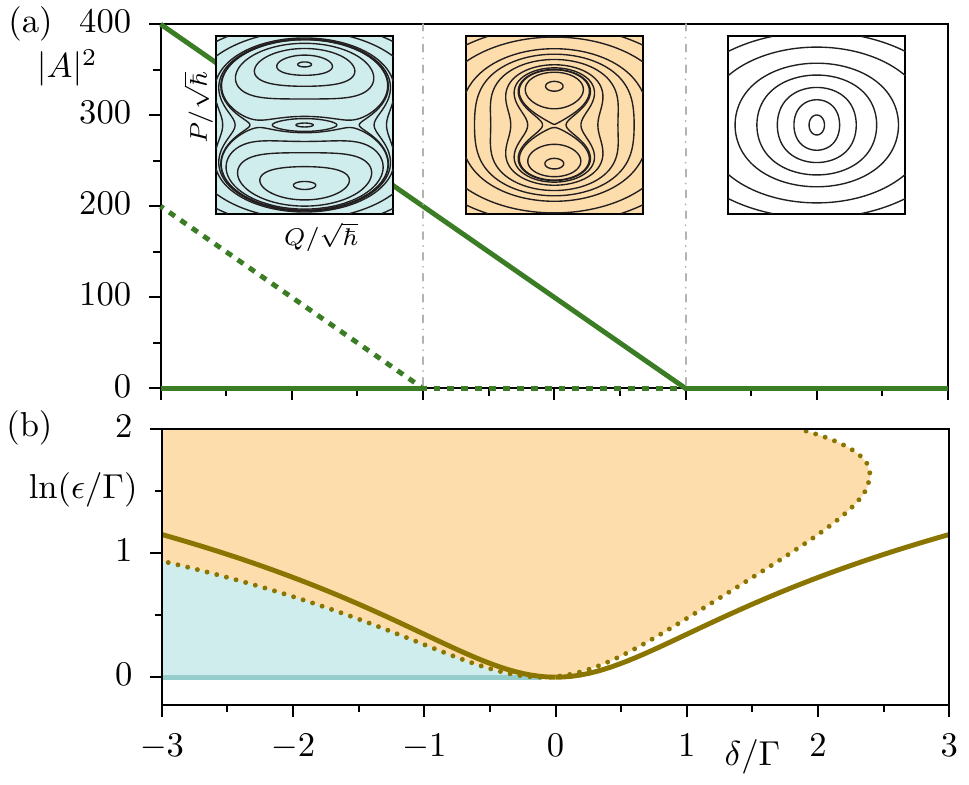}
\caption{
(a)
Amplitudes of cavity steady states vs.~detuning $\delta$ for $\epsilon = \sqrt{2} \Gamma$, stable (solid) and unstable (dotted); dashed vertical lines separate mono-, bi- and tristable regions; 
insets show  phase portraits of corresponding regions.  
(b)
Boundary of parametric instability without (solid) and with (dotted) account of nonlinear frequency shift, \Eq{eq:omega_shift}, for $\alpha=\Gamma/100$ and $\beta =1/10$, cf.~\Eq{eq:delta_max}; yellow region corresponds to bistable high-amplitude state, blue region indicates coexistence of stable high-amplitude and zero-amplitude states.
($\alpha=\Gamma/100$).
}
\label{fig:homogeneous}
\end{figure}

If $\epsilon < \Gamma$, only the trivial steady state, $A=0$, exists for all values of the detuning 
$\delta$.
If $\epsilon > \Gamma$, the trivial state turns unstable within
the interval $|\delta| \leq \dth$, and instead two non-trivial stable steady states, 
$A=|A| e^{\iexp \theta_A}$, emerge
at the threshold $\delta = \dth$, and persist for all $\delta < \dth$, 
see Fig.~\ref{fig:homogeneous}(a).
These states have identical amplitudes,
\begin{equation}\label{eq:a_hom_stable}
|A|^2 = {1\over \alpha}\left( -\delta + \sqrt{\epsilon^2 - \Gamma^2} \right)
\,,
\end{equation}
and are $\pi$-shifted in phase, with $\sin(2\theta_A) = \Gamma/\epsilon$.

In the further red detuned region, $\delta < -\dth$,  the trivial steady state solution, $A=0$,
becomes stable again, such that the three stable states coexist there. Simultaneously, 
two new unstable states emerge having the same amplitude,
$|A|^2 = \left( -\delta - \sqrt{\epsilon^2 - \Gamma^2} \right)/\alpha$. 

In the limit of $\epsilon=0$, $\Gamma=0$ (the undamped Duffing oscillator), the nontrivial stable and unstable states merge, forming a  manifold of marginally stable states with indefinite phase $\theta_A$ and amplitude $|A|^2 = -\delta/\alpha$.  

The steady states of the damped cavity at $\epsilon > \Gamma$  originate from the fixed points of the cavity Hamiltonian $H_{\text{cav}}$,  Eq.~\eqref{eq:Hcav_RF_QP}, which are illustrated in the insets of Fig.~\ref{fig:homogeneous}(a) for the mono-, bi-, and tristable regions.
The damping $\Gamma$  introduces the threshold for the emerging nontrivial states, and shifts the positions of the steady states in phase space away from the fixed points.

The pump parameters where new steady states occur
are determined by the stability properties of the underlying linear system,
characterized by the  matrix $\mathcal{A}(\alpha=0)$.
Its determinant, $D = \delta^2 + \Gamma^2 - \epsilon^2$,
causes divergence at the parametric instability threshold, $|\delta| = \sqrt{\epsilon^2 - \Gamma^2}$,
where the fixed point $A=0$ turns unstable. 
In a linear system this would lead to exponentially growing  solutions 
in the parameter regime $\epsilon>\Gamma$ and $|\delta| < \dth$,
with a rate $\lambda = -\Gamma + \sqrt{\epsilon^2 - \delta^2}$.
In the nonlinear system this global instability is lifted by the bifurcation of 
the fixed point $A=0$ into the two new stable steady states. 

The cavity field, as it leaks into the transmission line, generates an outgoing field
with the amplitude $C$ according to \Eq{eq:relation_ABC}.
For the steady state, \Eq{eq:a_hom_stable}, the flux radiated into the 
transmission line amounts to 
\begin{eqnarray}\label{eq:c_hom_stable}
 |C|^2 = {2\Gamma_0\over\alpha}  \left( -\delta + \sqrt{\epsilon^2 - \Gamma^2}\right)
\,.
\end{eqnarray}

The nonlinear effect of the cavity resonance shift  induced by the pump, mentioned in Sec.~\ref{sec:parametric}, \Eq{eq:omega_shift}, leads to the  quenching of the parametric instability at strong pumping as observed in experiment~\cite{IdaMaria_thesis}. 
By taking into account this shift, the actual pump detuning becomes 
$\Omega/2 - \omega_0(\epsilon) = \delta - (\omega_0\,(\epsilon) - \omega_0)$, and the parametric instability condition modifies accordingly,  
\begin{equation}\label{eq:delta_max}
\delta <  \sqrt{\epsilon^2-\Gamma^2} - {\beta\epsilon^2 \over \Gamma},
\end{equation}
where $\beta = \Gamma/\omega_0\gamma \tan^2{F} \ll 1$. The modified boundary of parametric instability in the $(\delta,\epsilon)$-plane is depicted in Fig.~\ref{fig:homogeneous}(b): 
the instability region is bounded by the maximum blue detuning,
$\delta_{\text{max}} = \Gamma/4\beta$,  
and  it is also bounded by a maximum pump strength at given detuning, 
e.g. $\epsilon_{\text{max}} = \Gamma/\beta$ at $\delta=0$.

\begin{figure}[tb]
\centering
\includegraphics[width=\columnwidth]{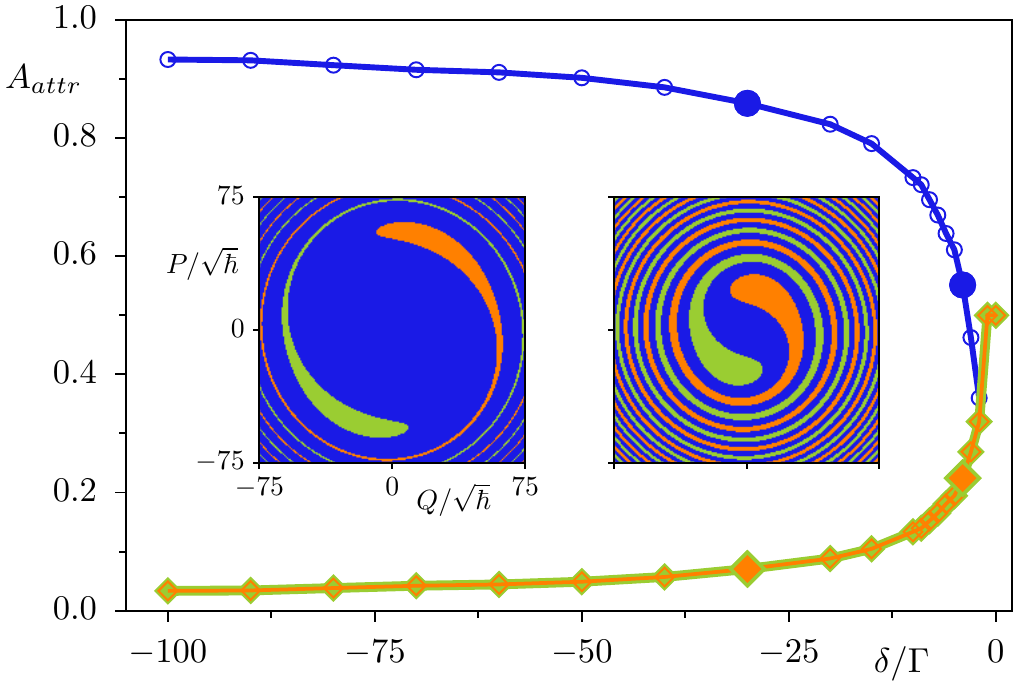}
\caption{
Relative areas of attractor basins of coexisting high-amplitude (lower curve) and zero-amplitude (upper curve) stable states vs.~detuning $\delta$. 
Insets show the basins of attraction of high-amplitude (red and green) and zero-amplitude (blue) states, for $\delta/\Gamma=-30$ (left) and $\delta/\Gamma=-4$ (right)
($\epsilon=\sqrt{2}\Gamma$, $\alpha=\Gamma/100$).
}
\label{fig:attr_basins}
\end{figure}

In the experiment~\cite{WilETAL2010}, all the described states of the parametrically pumped 
cavity have been observed: the subthreshold monostable regime at blue detuning, as well as the 
above-threshold bistable and tristable regimes at red detuning. The visibility of particular stable 
states in the multistable regime is defined by the probabilities of their occupation, which are 
determined by the relative areas of the respective basins of attraction, i.e.~the phase space 
regions from which trajectories asymptotically approach the respective state. 
Examples of the attractor basins in the red-detuned region, $\delta<-\dth$, are shown in the insets of Fig.~\ref{fig:attr_basins} where the blue basin belongs to the zero-amplitude state, 
and the red and green attractor basins are those of the high-amplitude states.
The relative areas of the latter rapidly decrease and become very small
in the far red-detuned region, as shown on the main panel in Fig.~\ref{fig:attr_basins}, implying that 
these states are much less populated. A similar conclusion is drawn from the calculation of 
the probability to escape from the high-amplitude states \cite{DykMalSmeSil1998}, which  is much larger than the one for the trivial state, $A=0$,  at far red detuning.

These arguments explain why in the experiment~\cite{WilETAL2010} the boundary of parametric resonance is washed out at red detuning, in contrast to the sharp boundary at blue detuning, which is determined by the threshold for the nontrivial steady states.

\subsection{Driven Duffing cavity ($\epsilon=0$)}
\label{sec:drivenduffing}

Now we turn to the discussion of the cavity response to a weak signal with zero detuning, 
$\Delta=0$, and complex amplitude $B = |B|e^{\iexp \theta_B}$.

\begin{figure}[tb]
\centering
\includegraphics[width=\columnwidth]{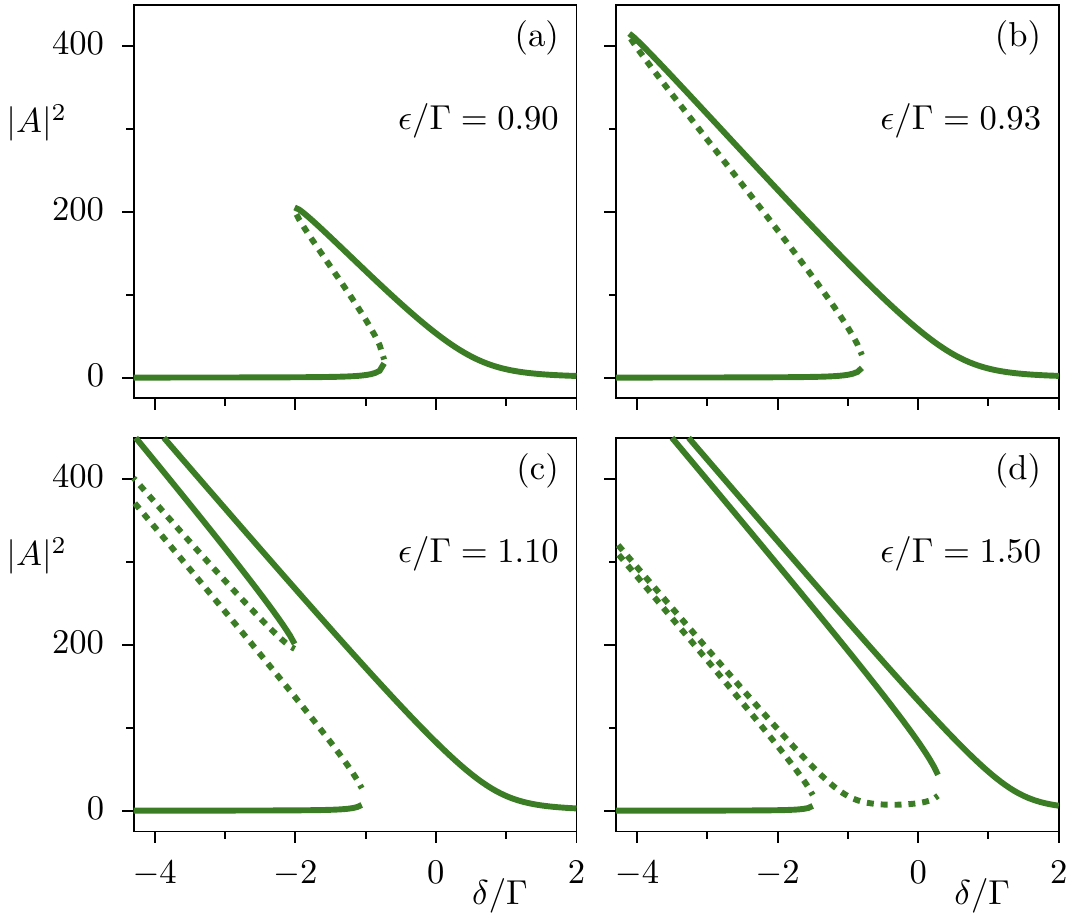}
\caption{
Steady state cavity response $|A|^2$ according to \Eq{eq:a_nonlin}
vs.~pump detuning $\delta$ for different values of the pump strength $\epsilon$,
below threshold, $\epsilon/\Gamma = 0.9, 0.93$ (a,b), and above threshold,  
$\epsilon = 1.1, 1.5$ (c,d).
Solid and dashed lines mark stable and instable states, respectively.
($|B|^2 = 2 \Gamma$, $\theta_B = \pi/2$, $\alpha = \Gamma/100$, $\Gamma_R = 0$).
}
\label{fig:AsqRsq_delta}
\end{figure}

\begin{figure}[tb]
\centering
\includegraphics[width=\columnwidth]{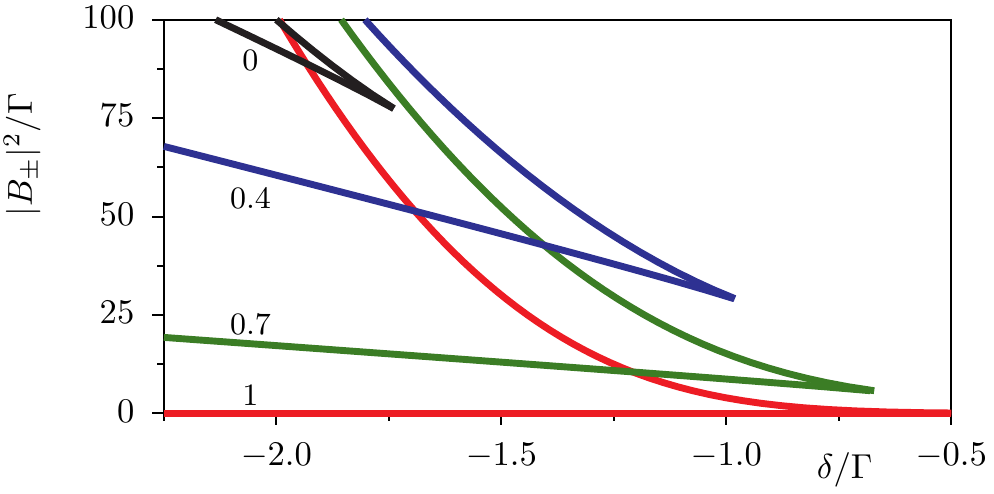}
\caption{
Bistability regions of the cavity response, bounded by bifurcation lines $|B_\pm(\delta)|^2$,
for different subthreshold values of the pump strength,
$\epsilon/\Gamma = 0, 0.4, 0.7, 1.0$  
($\theta_B = \pi/2$, $\alpha = \Gamma/100$, $\Gamma_R = 0$).
}
\label{fig:dpDbifurc_Ain}
\end{figure}

It is instructive to first review the response of the driven Duffing oscillator~\cite{Nayfeh}, which corresponds to the limit $\epsilon = 0$ in \Eq{eq:EOM_a_matrix}. In this case the detuning 
$\delta$ refers to the deviation of the input frequency from the cavity resonance. The cavity response is given by the equation, 
\begin{equation}
A = {\sqrt{2\Gamma_0} \over \zeta + i \Gamma }\, B 
\,.
\end{equation}
The maximum response is achieved at $\zeta=0$, along the tilted line 
$|A|^2(\delta)=-\delta/\alpha$,
and amounts to $|A_{\text{max}}|^2 = |A|^2(\zeta=0) = 2 \Gamma_0 |B|^2/\Gamma^2$, independent of $\alpha$.
As a consequence of the tilted  resonance line, the cavity response can display bistability, with two coexisting stable states, as shown in Fig.~\ref{fig:AsqRsq_delta}(a). The bistability emerges above the critical  value of the driving amplitude, $|B_c|^2 = 4 \Gamma^3/(3 \sqrt{3} \alpha \Gamma_0)$, and at the detunings, $\delta<\delta_{c} = -\sqrt{3}\Gamma$. The bistability region is confined by the bifurcation lines,
\begin{equation}
|B_\pm | = {\delta^3\over 27 \alpha \Gamma_0} 
\left[-1- {9\Gamma^2\over \delta^2} \pm \left(1- {3\Gamma^2\over \delta^2}\right)^{3/2} \right]
\,,
\end{equation}
forming a wedge in the ($\delta$-$|B|^2$) plane, as illustrated in Fig.~\ref{fig:dpDbifurc_Ain} with black lines.

An ideal Duffing cavity fully reflects the  input signal, so the amplitude of the output,  $C=|C|e^{\iexp \theta_C}$, carries no information about the resonance, $|C|=|B|$. Such information is only available for a lossy cavity, where
\begin{equation}\label{eq:duffingrefl}
{|C|^2\over|B|^2} = 1  - {4\Gamma_0\Gamma_R\over \Gamma^2 + \zeta^2}
\,.
\end{equation}
On the other hand, the phase $\theta_C$ of the output signal is sensitive to the position of the resonance. This is the working principle of the Josephson bifurcation amplifiers
\cite{JBAreview2009},
where the variation of $\theta_C$ under sweeping the input power through the bistability region
is exploited for the qubit readout.

\subsection{Driven parametric cavity}
\label{sec:drivenparam}

Switching on the parametric pumping, $\epsilon > 0$, qualitatively changes the cavity response. Now the amplitude of the cavity field is determined by the equation,
\begin{eqnarray}\label{eq:a_nonlin}
 \displaystyle {|A|^2\over |B|^2} &=& {2 \Gamma_0 \over D^2}
\left(
\zeta^2 + \Gamma^2 + \epsilon^2  
- 2 \epsilon [\zeta \cos{2\theta_B} 
+  \Gamma \sin{2\theta_B}] 
\right)
, \nonumber\\
\label{eq:detD}
D &=& \det(\mathcal{A}) = \zeta^2 + \Gamma^2 - \epsilon^2 
\,.
\end{eqnarray}
%
In the subthreshold regime $\epsilon < \Gamma$, the cavity response  
remains qualitatively similar to the Duffing oscillator, 
see Figs.~\ref{fig:AsqRsq_delta}(a)-(b).
The role of the parametric pumping in this regime is to effectively reduce the damping term, 
$\Gamma^2 \to \Gamma^2 - \epsilon^2$. This makes the resonance more narrow and, at the same time, strongly increases the cavity amplitude along the tilted resonance line 
$\zeta=0$. Another important feature is an explicit dependence of the cavity field on the phase shift $\theta_B$ of the input with respect to the parametric pump.  

The maximum value of the cavity field is,
\begin{equation}
|A|^2(\theta_B, \zeta=0) 
= 2 \Gamma_0  
\frac{\Gamma^2 + \epsilon^2 - 2 \epsilon \Gamma \sin(2 \theta_B)}{\left(\Gamma^2 - \epsilon^2 \right)^2}|B|^2
\,.
\end{equation}
Similar to  the Duffing limit, this value is independent of the nonlinearity coefficient $\alpha$. The maximum response diverges at  $\epsilon=\Gamma$, which can be compared to the resonance 
catastrophe of a linear parametric oscillator.
While in the linear case the divergence occurs at $\delta = \pm \sqrt{\epsilon^2 + \Gamma^2}$, the nonlinearity here shifts the divergence towards an infinite red detuning. 

As a consequence of the resonance narrowing, the 
critical bifurcation point moves towards the origin, $|B_c|^2=\delta_c =0$ when $\epsilon \to 
\Gamma$, as illustrated in Fig.~\ref{fig:dpDbifurc_Ain}. 

Above the threshold, $\epsilon > \Gamma$, the resonance splits into two branches, as 
shown in Figs.~\ref{fig:AsqRsq_delta}(c)-(d), each branch consisting of two non-degenerate 
steady states, one pair being stable and the other unstable. These states originate from the degenerate nontrivial states in the absence of an input signal, cf.~Fig.~\ref{fig:homogeneous}, the degeneracy being now lifted by the input. The distance between the branches increases with $\epsilon$. 

\begin{figure}[tb]
\centering
\includegraphics[width=\columnwidth]{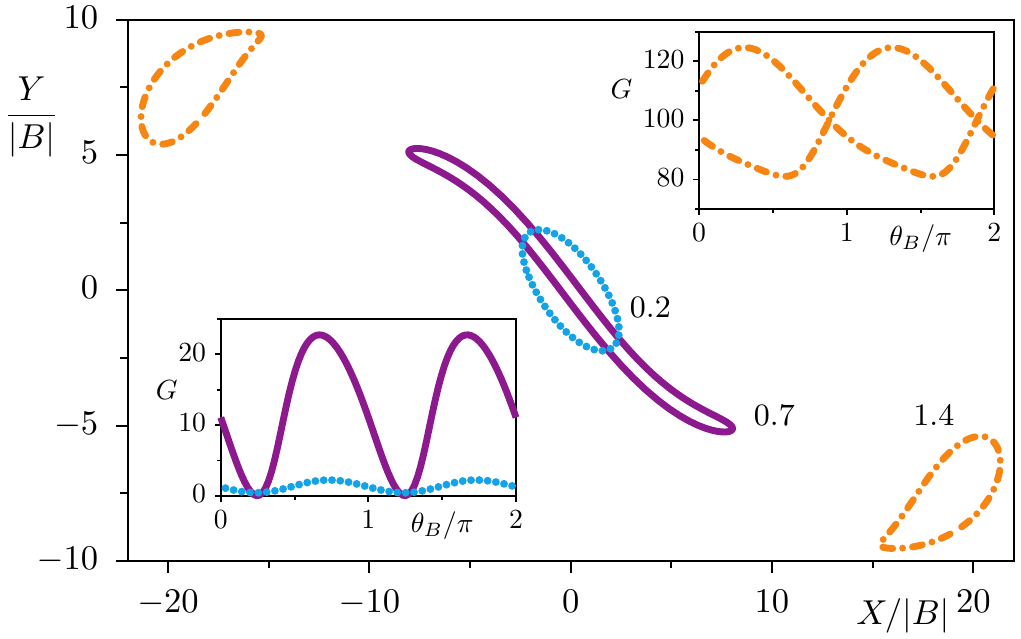}
\caption{
Anisotropy of the cavity output field in the complex $C$-plane, in the subthreshold regime ($\epsilon/\Gamma = 0.2, 0.7$ [blue, red]) and above the threshold 
($\epsilon/\Gamma=1.4$ [yellow]); 
insets show the dispersion of the gain with the input phase $\theta_B$ below threshold (left) and above (right).
($\delta = 0$, $|B|^2=2\Gamma$, $\alpha = \Gamma/100$, $\Gamma_R = 0$).
}
\label{fig:Rsq_phin}
\end{figure}

The scattering by the parametrically pumped cavity is always inelastic, in contrast to the Duffing cavity,  and the output signal in general differs significantly from the input signal, 
not only in phase but also in the absolute value, $|C| \neq |B|$.
Using the input-output relation in \Eq{eq:relation_ABC}, and the steady state solution $|A|^2$ in 
\Eq{eq:EOM_a_matrix}, 
the output amplitude can be expressed as a function of the input amplitude,
\begin{eqnarray}\label{eq:relation_v_vin}
&&\left( \begin{array}{l} C \\ C^{*}   \end{array} \right)
= 
{\cal V}
\left( \begin{array}{l} B \\ B^{*}   \end{array} \right) \,, \nonumber\\
&&{\cal V} = 
\left(\begin{array}{cc}
\sqrt{1+q^2 - q_R} e^{\iexp \eta} & \ui q \\
-\ui q & \sqrt{1+q^2 - q_R} e^{-\iexp \eta}
\end{array}\right)
,
\end{eqnarray}
with the parameters
\begin{eqnarray}
\label{eq:def_q}
q &=& {2\epsilon \Gamma_0 \over D}\,,  \quad 
q_R = {4\Gamma_0\Gamma_R\over D} \\
\label{eq:def_eta}
\eta &=& \arctan\left(
\frac{-2\Gamma_0\zeta}{\zeta^2-\Gamma_0^2 + \Gamma_R^2 - \epsilon^2}
\right)
\,.
\end{eqnarray}
The relation in \Eq{eq:relation_v_vin} maps the points of the unit circle,
$B = e^{\iexp \theta_B}$, 
onto the phase-dependent curve, $C(\theta_B)$, and determines the phase-dependent gain
\begin{eqnarray}\label{eq:c2_b2} 
&&G(\theta_B) = {|C(\theta_B)|^2 \over |B|^2} \nonumber\\
&& =  1 + 2q^2  - q_R + 2q \sqrt{1+q^2 - q_R} \sin\left(2\theta_B + \eta\right) 
\,.
\end{eqnarray}
The $\theta_B$-dependence of the gain $|C|^2/|B|^2$ and the  quadratures of $C=(X+\ui Y)/2$
are illustrated in Fig.~\ref{fig:Rsq_phin}.
In the monostable (subthreshold) regime the output amplitude $C$ is amplified ($G>1$) 
or deamplified ($G<1$) depending on the input phase. 
For $\epsilon \lesssim \Gamma$ the points $C(\theta_B)$ form a strongly elongated curve 
in phase space, centered at $(0,0)$.
In the quasilinear regime, where the parameters $q$ and $q_R$ in \Eq{eq:c2_b2}
are approximately independent of $\theta_B$, this curve approaches an ellipse with the half axes
\begin{eqnarray}\label{eq:gain12}
 \sqrt{G_{max,min}} = \sqrt{1+q^2 - q_R} \pm q
\,,
\end{eqnarray}
giving the maximum / minimum gain factor along those quadratures.
For negligible internal losses, $q_R/q^2 \ll 1$,
the amplified and deamplified quadratures are related according to $G_{\text{min}} = 1/G_{\text{max}}$.

In the limit $\epsilon \to 0$  the gain factors become equal and 
reduce to the reflection coefficient of the Duffing oscillator, \Eq{eq:duffingrefl}.

In the bistable regime above the threshold the corresponding output amplitudes $C(\theta_B)$ are mapped on two distinct closed curves in phase space, with a $\pi$-phase shift  between them, as shown in  Fig.~\ref{fig:Rsq_phin}. The offset from the origin is due to the parametric radiation generated by the cavity.

%

\subsection{Response to detuned signal}
\label{subsec:detuned}

\begin{figure}[tb]
\centering 
\includegraphics[width=\columnwidth]{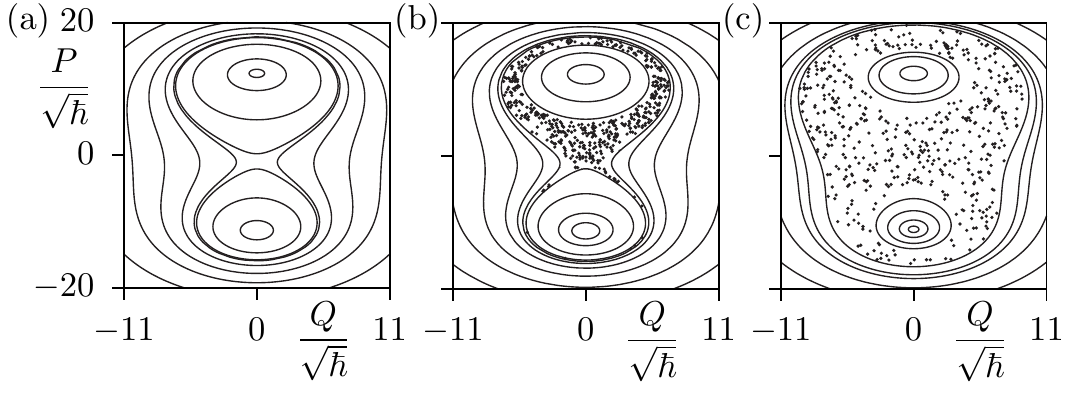}
\caption{
Phase-space representation of cavity amplitude $A = (Q + \ui P)/\sqrt{2\hbar}$ 
driven by detuned input amplitude $B(t) = B e^{-\iexp \Delta t}$. 
The trajectories $(Q(t),P(t))$ are evaluated in the conservative limit, 
neglecting the damping term on the lhs of \Eq{eq:EOM_A_cl}, 
for (a) $\Delta = 0$, (b) $\Delta = \Gamma/10$, and (c) $\Delta = \Gamma$.
In (b)-(c) trajectories are represented stroboscopically,
at times $t = 2\pi n/\Delta$, ($n=0,1,2,\ldots$).
($\epsilon = \sqrt{2}\Gamma$, $\delta=0$, $|B|^2 = \Gamma$, $\theta_B=\pi/2$, $\alpha=\Gamma/100$, $\Gamma_R = 0$). 
}
\label{fig:phasespace}
\end{figure}

The cavity response has a simple stationary form only when the frequency $\omega$ of the input signal strictly matches the half-frequency of the pump, $\Omega/2$. If the input is time-dependent in the rotating frame, e.g. $B(t) = B e^{-\iexp \Delta t}$ with $\Delta = \omega-\Omega/2$, the combination of the time-periodic force with the nonlinearity leads to the formation of a region in phase space where the cavity amplitude $A$ evolves chaotically, as illustrated in Fig.~\ref{fig:phasespace} for the bistable regime above the parametric threshold.
With increasing input amplitude and detuning a chaotic layer forms around the instable fixed points of the Hamiltonian~\eqref{eq:Hcav_RF_QP_B}, and its area grows with $|B|$ and 
$|\Delta|$, see Fig.~\ref{fig:phasespace}. However, as long as the stable fixed points persist in the presence of the time-dependent drive, the amplitude of the damped cavity 
evolves into a time-periodic limit cycle around them,
and then the time-average of $A(t)$ gives only small corrections to the stationary result.

In this section we evaluate the response of an ideal cavity $(\Gamma_R=0)$  to a detuned signal in the monostable regime, $\epsilon < \sqrt{\delta^2 + \Gamma^2}$. We restrict to a linear response assuming  $\alpha |A|^2\ll \sqrt{\Gamma^2 + \delta^2 - \epsilon^2}$. 

Suppose the input signal in \Eq{eq:EOM_A_cl} consists of two conjugated harmonics, 
$B(t)= B(\Delta) e^{-\iexp\Delta t} + B(-\Delta) e^{\iexp\Delta t}$ 
(signal and idler in the terminology of non-degenerate parametric amplification). 
Then the  output field, as well as the field in the cavity, will also consist of the combination of the same harmonics. The output amplitudes are related to the input via the equation generalizing 
\Eqs{eq:relation_v_vin}-(\ref{eq:def_q}),
\begin{eqnarray}\label{eq:relation_v_Delta}
\left( \begin{array}{l} C(\Delta) \\ C^{*}(-\Delta)   \end{array} \right)
= 
 {\cal V}(\Delta)
\left( \begin{array}{l} B(\Delta) \\ B^{*}(-\Delta)  
 \end{array} \right) \,, 
\end{eqnarray}
where
 \begin{eqnarray}\label{eq:v_ik}
 {\cal V}(\Delta) &=& {1\over D(\Delta)}
\left(\begin{array}{cc}
v_{11}(\Delta) & v_{12} \\
v_{12}^\ast & v_{11}^\ast (\Delta)
\end{array}
\right)\,,\\
v_{11}(\Delta) &=&   (\delta - \ui \Gamma)^2-\Delta^2 - \epsilon^2, \;
v_{12} =  2 \ui \Gamma\epsilon\,, \nonumber\\
\label{eq:D(Delta)}
D(\Delta) &=&  (\Gamma -\ui \Delta)^2 + \delta^2 - \epsilon^2\,.
\end{eqnarray}
The coupling between the conjugated harmonics is a fingerprint of parametric amplification: an input at frequency $\Delta$ generates outputs at frequencies $\Delta$ and $-\Delta$, and conversely an output at frequency $\Delta$ consists of the contributions of inputs at frequencies $\Delta$ and  $-\Delta$. In particular,  for  $B(-\Delta)=0$, \Eq{eq:relation_v_Delta} yields,
\begin{eqnarray}\label{eq:Cw_Bw_cl}
 &C& (\Delta) = \sqrt {1+ |q(\Delta)|^2} e^{\iexp \eta(\Delta)} B(\Delta)\,, 
\ \nonumber \\
&C&(-\Delta) =  \ui q^\ast(\Delta) B^\ast(\Delta) \,,\\
\label{eq:qw_cl}
&q& (\Delta) = {2\epsilon \Gamma_0 / D(\Delta)}
\,.
\end{eqnarray}
%

Amplification of the detuned signal is characterized by two gain factors, direct gain 
$G_1(\Delta) = |C(\Delta)|^2/|B(\Delta)|^2 = 1 + |q(\Delta)|^2$, and interconversion gain,  
$G_2(\Delta) = |C(-\Delta)|^2/|B(\Delta)|^2 =|q(\Delta)|^2 $.  These two gains are fundamentally related, $G_2=G_1 - 1$,  which is the consequence of the fundamental property of the matrix elements in \Eq{eq:v_ik}, $|v_{11}|^2 - |v_{12}|^2 = |D(\Delta)|^2$. For the quantum fields, this property  guarantees the unitary relation between the input and output quantum states (see later in Sec.~\ref{sec:quantum}).  

\begin{figure}
\centering 
\includegraphics[width=\columnwidth]{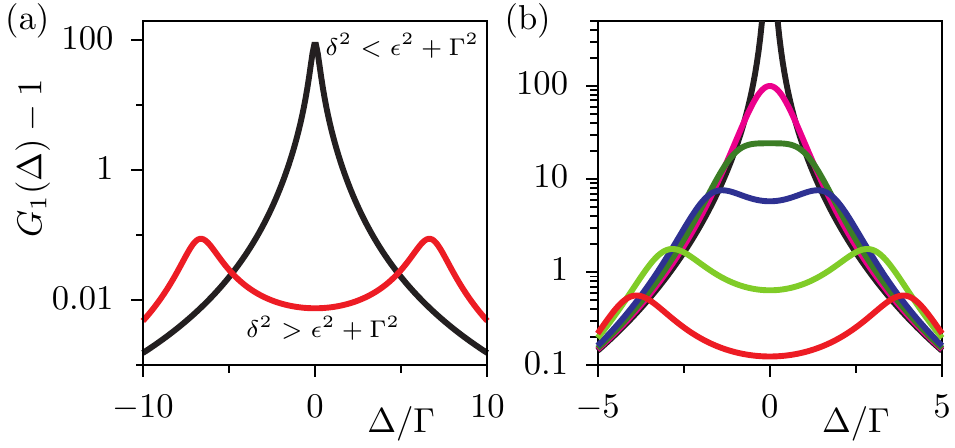}
\caption{
Linear gain of detuned input signal; 
(a) $G(\Delta)$ for small detuning $\delta = 1.85\Gamma$ (single resonance), and large detuning $\delta = 7\Gamma$ (split resonance), at fixed pump strength 
$\epsilon/\Gamma = 2$;   
(b) $G(\Delta)$ for fixed detuning $\delta = 5\Gamma$ and increasing pump strengths, 
$\epsilon/\Gamma = 3 \ldots \sqrt{1+\delta^2/\Gamma^2}$ (from bottom to top).
($\Gamma_R = 0$).
}
\label{fig:gain_dw__dpD}
\end{figure}

The amplification of detuned signals possesses another interesting property - the appearance of resonance features, as illustrated in Fig.~\ref{fig:gain_dw__dpD}(a).
The resonance structure of the gain is determined by the determinant $|D(\Delta)|^2$, \Eq{eq:qw_cl}. 
 It has a single minimum, at $\Delta = 0$, within the interval of relatively small detuning, 
 $\Gamma^2 - \epsilon^2<\delta^2 < \epsilon^2 + \Gamma^2$, and the gain factor $G(\Delta)$ is accordingly single peaked at $\Delta=0$.
However, at larger detunings,
\begin{equation}\label{eq:cond_twores_cl}
\delta^2 > \epsilon^2 + \Gamma^2,  
\end{equation}
 two resonance peaks emerge, situated symmetrically with respect to $\Delta=0$ at
\begin{equation}\label{eq:wres_cl}
\Delta = \pm \sqrt{\delta^2 - \epsilon^2- \Gamma^2 }
\,.
\end{equation}
The origin of these resonances can be understood from the behavior of the response function of a conventional damped  linear oscillator, 
$\chi(\omega) = (\omega_0^2 - \omega^2 - \ui\omega\Gamma)^{-1}$. 
At small damping, $\Gamma \ll \omega_0$, the resonance is close to the eigen frequency 
$\omega_0$, $\omega=\sqrt{\omega_0^2-\Gamma^2/2}$. With increasing damping the resonance is pulled towards the zero frequency, and stays at the zero frequency as soon as   
$\Gamma>\sqrt{2}\omega_0$. 
Similarly, the resonances in the response of the linearized parametric oscillator, 
\Eq{eq:EOM_A_cl} with $\alpha=0$, are at small $\Gamma$ close to the oscillator eigen frequencies, 
$\pm \sqrt{\delta^2 - \epsilon^2}$, as in \Eq{eq:wres_cl}, but are pulled towards $\Delta=0$ with increasing $\Gamma$, and eventually merge when $\Gamma > \sqrt{\delta^2 - \epsilon^2}$, \Eq{eq:cond_twores_cl}.

\section{Amplification and qubit readout}
\label{sec:application}

In this section we discuss the application of the parametrically pumped cavity for signal amplification, and for dispersive qubit  readout. 

In what follows we shall neglect internal losses in the cavity and assume 
$\Gamma_0=\Gamma$.

\subsection{Amplification}
\label{sec:amplification}

The amplification characteristics of the nonlinear parametric cavity depend on many parameters: pump and input strengths and detunings from the cavity resonance, relative phase shift, nonlinearity and damping, which makes the overall picture pretty intricate. 

\begin{figure}[bt]
\centering
\includegraphics[width=\columnwidth]{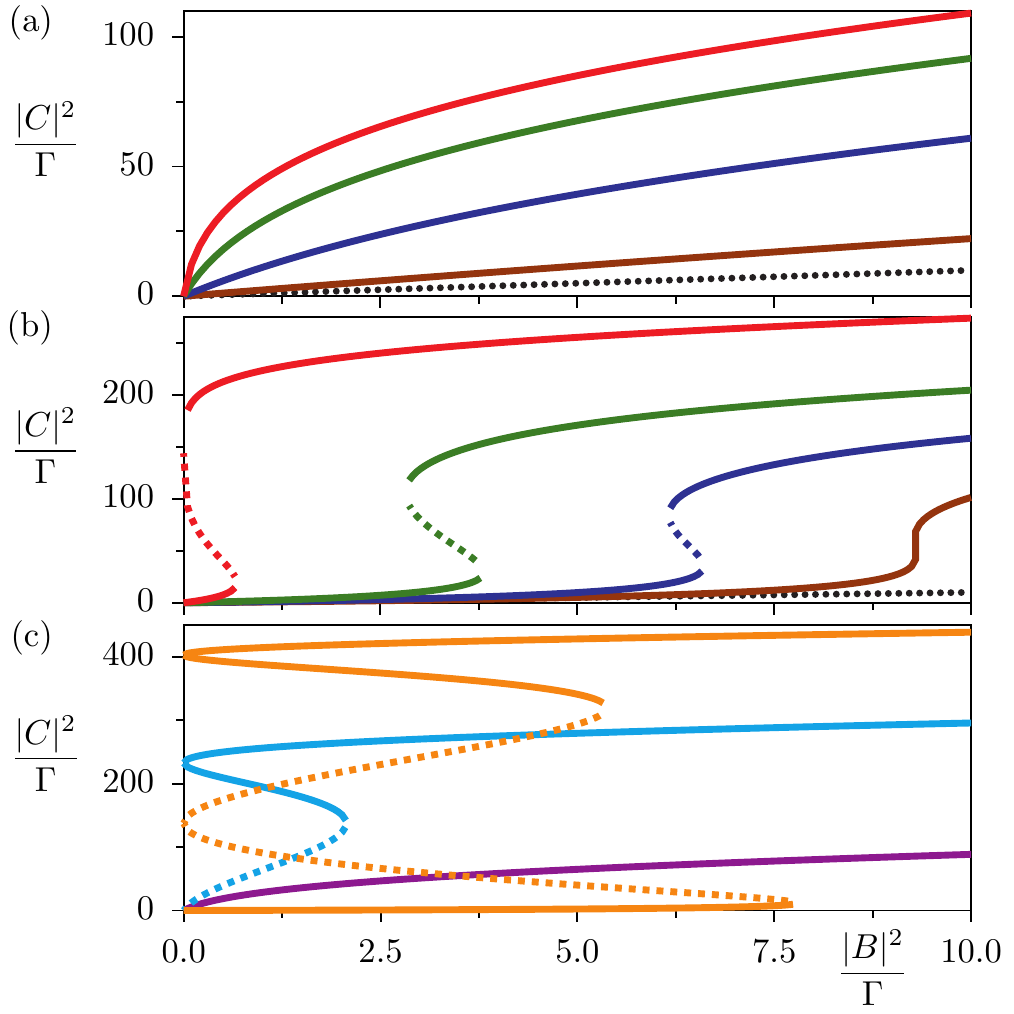}
\caption{
Output power $|C|^2$ vs.~input power $|B|^2$,
below threshold, $\epsilon < \Gamma$ (a,b), 
and above threshold, $\epsilon = 1.2\Gamma$ (c).
(a) $\delta/\Gamma=0.5$ 
and $\epsilon/\Gamma = 0, 0.3, 0.7, 0.9, 1.0$ (from bottom to top);
(b) $\delta/\Gamma=-0.72$ 
and $\epsilon/\Gamma = 0, 0.63, 0.7, 0.8, 1.0$ (from bottom to top);
the dotted line refers to the Duffing limit, $\epsilon=0$.
(c) $\epsilon/\Gamma=1.2$ and 
$\delta/\Gamma = -1.4, -0.5, 1.0$ (from top to bottom)
For each of the parameters instable branches are indicated by dashed lines.
($\theta_B=\pi/2$, $\alpha=\Gamma/100$).
}
\label{fig:dRsqdBsq_Ain}
\end{figure}

 The output power $|C|^2$ as a function of the input power 
$|B|^2$ for on-resonance input, $\Delta=0$, is depicted in Fig.~\ref{fig:dRsqdBsq_Ain} for various values of pump strengths and pump detunings. The major phenomenon here is the appearance of multistable regimes.  The bistable regime establishes already below the threshold, $\epsilon<\Gamma$, in the red detuning region, $\delta<0$, as shown on Fig.~\ref{fig:dRsqdBsq_Ain}(b). Above the threshold, the 
mono-, bi-, and tristable regimes exist at different detunings, as shown on 
Fig.~\ref{fig:dRsqdBsq_Ain}(c). Moreover, in the latter regime, the output power does not approach zero value at $|B|^2 = 0$ due to the effect of parametric radiation.   

For the amplification purpose the monostable regime in Fig.~\ref{fig:dRsqdBsq_Ain}(a) is the most suitable. The output power in this regime depends monotonically on the input power, but exhibits pronounced nonlinearity with increasing pump strength at input power levels 
$|B|^2\sim\Gamma$. 

The maximum differential gain is achieved at small input power, and for phase shift $\theta_B \approx -\pi/4$. 
The gain is controlled by the quantity $q$ in \Eq{eq:gain12}, and at large $q \gg 1,\, q_R$,
\begin{eqnarray}
{G_{\text{max}}} 
\approx (2q)^2 = \left({4\epsilon\Gamma\over D} \right)^2\gg 1
\,.
\end{eqnarray}
The gain increases while approaching the threshold (cf.~Fig.~\ref{fig:AsqRsq_delta}(b)), $G_{\text{max}} \approx (4\epsilon\Gamma)^2/(\delta^2+\Gamma^2 - \epsilon^2)^2$, in the quasilinear approximation, and then it is limited by the nonlinearity. 
Let us evaluate this upper bound for the gain at $\epsilon=\Gamma$ and $\delta=0$. In this case, 
$G_{\text{max}} = (2\Gamma/\alpha|A|^2)^4$. Extracting the amplitude $|A|^2$ from 
\Eq{eq:a_nonlin}, with $\theta_B=-\pi/4$,
\begin{equation}\label{Amax}
\alpha|A|^2 \approx \Gamma \left({8\alpha |B|^2 / \Gamma^2}\right)^{1/5},
\end{equation}
we get
\begin{equation}\label{Gmax_nonl}
G_{\text{max}} = \left(4\Gamma / \alpha\right)^{4/5} \left(|B|^2 / \Gamma\right)^{-4/5}\,.
\end{equation}

In a similar way we can evaluate the absolute minimum of deamplification. 
This is achieved at $\theta_B\approx \pi/4$, where $G_{\text{min}}\approx 1/(2q)^2$, and 
\begin{equation}\label{Amin}
\alpha|A|^2 \approx \Gamma \left({2\alpha |B|^2 / \Gamma^2}\right)^{1/3},
\end{equation}
leading to the equation for minimum gain,
\begin{equation}\label{Gmin_nonl}
G_{\text{min}} = \left(\alpha/4\Gamma \right)^{4/3} \left(|B|^2 / \Gamma\right)^{4/3}\,.
\end{equation}

We note that the nonlinear deamplification is more efficient than the amplification: the product of the maximum and minimum nonlinear gains significantly deviates from unity, in contrast to the linear case, 
\begin{equation}\label{Gproduct}
G_{\text{max}}G_{\text{min}} \approx \sqrt{\alpha|B|^2 / 4\Gamma^2} < 1.
 \end{equation}
With these results we conclude that the maximum amplification (deamplification) efficiency is controlled by the parameter 
$\Gamma/\alpha$, and therefore a relatively small nonlinearity coefficient is required to achieve a large parametric effect. 

As we will see later, the same conclusion is also valid for the nonclassical properties of the fluctuations. 

At this point it is appropriate to estimate the output signal-to-noise ratio for parametric amplification, referring to the results of the noise analysis in Sec.~\ref{sec:noiseC}. According to \Eqs{eq:nc_vac_tot} and \eqref{eq:Csq_quasilin} the amplified noise increases in the vicinity of the threshold, however, the noise amplification is less efficient than the signal amplification, giving the ratio (for the quasilinear limit),
\begin{equation}
{|C|^2\over \ncvac} \approx {8\over \Gamma-\epsilon}\, |B|^2
\,.
\end{equation}
This ratio is large as soon as $|B|^2 > (\Gamma-\epsilon)/8$.

Amplification of a detuned signal, $\Delta\neq 0$, has qualitatively similar properties in the vicinity of the parametric threshold, $\epsilon^2 - \delta^2  \lesssim  \Gamma^2$. 
Here the gain factor $G(\Delta)$ has a quasi-Lorentzian shape, peaked at $\Delta=0$, as shown in 
Fig.~\ref{fig:gain_dw__dpD}(b),  the maximum gain increases while approaching the parametric threshold, while the bandwidth shrinks to zero.

However, the  bandwidth can be considerably increased, maintaining rather high gain, by working away from the parametric threshold in the region where the gain peak splits, $ \delta^2 \sim \epsilon^2 + \Gamma^2$, \Eq{eq:cond_twores_cl}. Here a wide frequency plateau emerges around $\Delta = 0$, see Fig.~\ref{fig:gain_dw__dpD}(b), where the gain factor is nearly constant over a frequency interval given by the distance between the resonances,   
$\Delta = \pm \sqrt{\delta^2 - \epsilon^2 - \Gamma^2}$.

\subsection{Bifurcation readout below threshold}
\label{sec:JBA}

\begin{figure}[tb]
\includegraphics[width=\columnwidth]{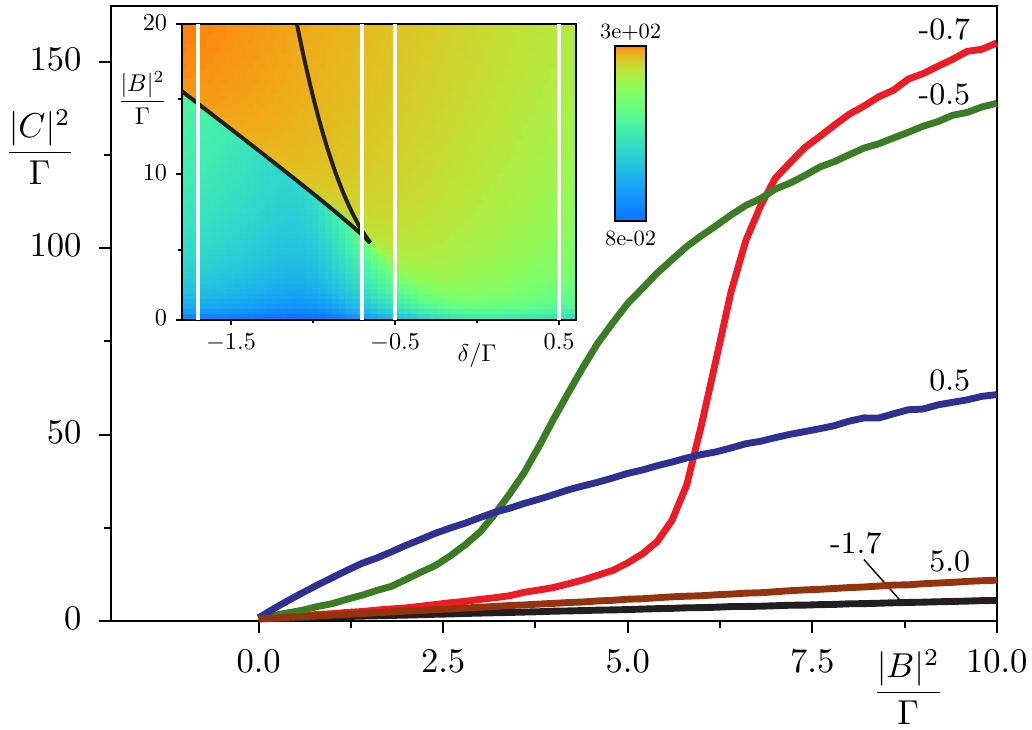}
\caption{
Average output power $\langle |C|^2 \rangle$ vs.~input power $|B|^2$
for $\epsilon/\Gamma = 0.7$
and $\delta/\Gamma = -1.7, -0.7, -0.5, 0.5, 5.0$. 
Inset: maximum gain factor G vs.~$|B|^2$ and detuning $\delta$;
black lines indicate the boundaries of the bistable region (wedge) 
cf. Fig.~\ref{fig:dpDbifurc_Ain}; white vertical lines indicate the parameter traces used in the main figure. 
The average output power $\langle |C|^2 \rangle$ is obtained from \Eq{eq:EOM_A_cl} in the presence of white Gaussian noise in the input.
($\theta_B=\pi/2$, $\alpha=\Gamma/100$).
}
\label{fig:meanRsq_Ain}
\end{figure}

The bifurcation regime of the cavity nonlinear response in the absence of parametric pumping is employed for dispersive qubit readout using JBA \cite{SidETAL2004}, for a review see Ref.~\onlinecite{JBAreview2009} and references therein. With this method, the phase shift of a reflected (or transmitted) probing signal is measured while ramping the signal amplitude.
The result is sensitive to the detuning of the signal tone from the cavity resonance, which is pulled by the qubit by $\pm\delta_q$, depending on the qubit state.  

One may take advantage of the high parametric gain for probing a qubit state by measuring the {\em amplitude} of the output signal instead of the phase shift. The amplified signal exhibits significant dispersion over the cavity-pump detuning  thus providing high contrast for the qubit readout. 

The basis of the method can be understood from Fig.~\ref{fig:meanRsq_Ain}; here the average output power is plotted against the input power for different detunings below the threshold, $\epsilon = 0.7\,\Gamma$. 
The bistability wedge for this pump strength is illustrated in the inset, compare also Fig.~\ref{fig:dpDbifurc_Ain}. 
The lowest three curves in Fig.~\ref{fig:meanRsq_Ain} correspond to values of the detuning within the monostable regions, either to the right or to the left of the critical bifurcation point, as indicated by white cuts in the inset (in the latter case, $\delta = -1.7\,\Gamma$,  the ramped input signal should not cross the bifurcation line).
The other two curves correspond to crossing through the bistability wedge or very close to the critical bifurcation point (here the average output power in the presence of classical noise is plotted, which then exhibits a gradual transition from the low- to the high-amplitude branch of the bifurcation curve).

The output contrast is extremely sensitive to the detuning: it is up to factor of 10 for detunings differing by  a  linewidth $\Gamma$ already at rather small input power, $|B|^2 \sim 10\,\Gamma$. 
In practice, a cavity frequency pull exerted by the qubit may be of the order~\cite{WalETAL2007} 
$\delta_q \sim 10 \uMHz$, i.~e.~comparable to the linewidth, $\Gamma \approx 10^{-4} \omega_0 \lesssim 10 \uMHz$. 

The output contrast can be further enhanced by increasing the pump strength towards the threshold. It is also possible to ramp the pump strength rather than input power. The possibility to operate with several parameters gives room for further optimization.

\subsection{Radiation readout above threshold}
\label{sec:qubit}

\begin{figure}[tb]
\centering
\includegraphics[width=\columnwidth]{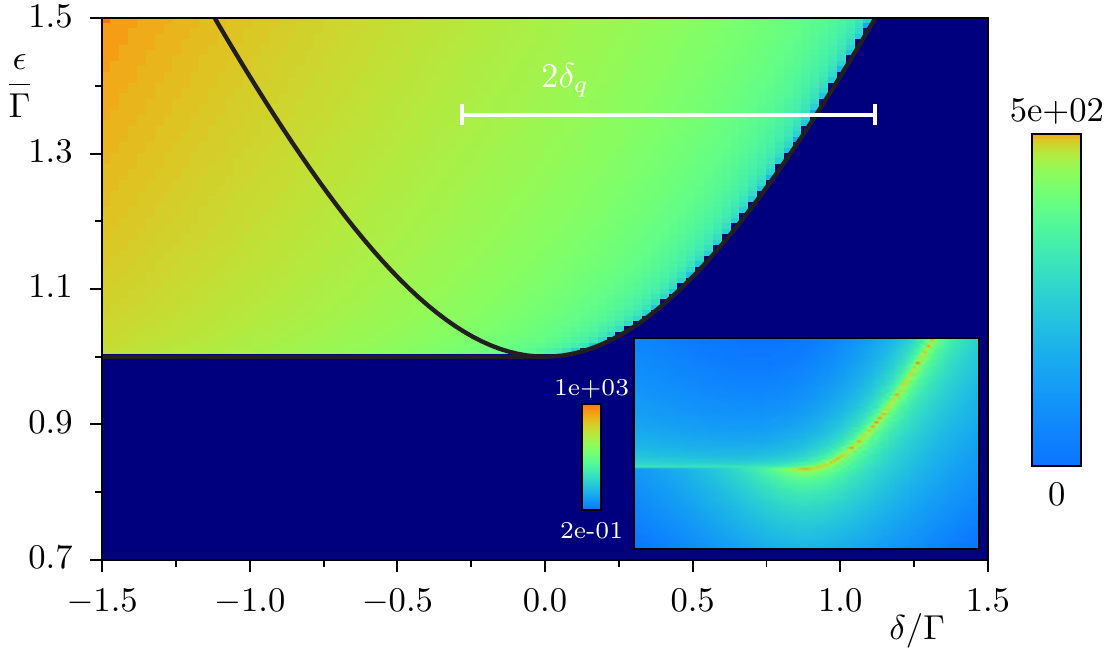}
\caption{
Output power $|C|^2/\Gamma$ 
vs.~detuning $\delta$ and pump strength $\epsilon$ for $B = 0$.
The black line separates bistable and tristable regimes, cf.~Fig.~\ref{fig:homogeneous};
the white line spans between two qubit-state dependent, effective detunings 
$\delta_0 \pm \delta_q$ 
($\delta_q = 0.7\Gamma$, 
$\epsilon > \sqrt{\Gamma^2 + \delta_q^2}$, 
$\delta_0 \gtrsim \dth - \delta_q$,
$\alpha=\Gamma/100$).
Inset: noise photon number $\ncvac/\Gamma$, \Eq{eq:nc_vac_tot}, vs.~$\delta$, $\epsilon$ as in main figure. 
}
\label{fig:Rsq_delta_epsilon}
\end{figure}

An alternative strategy for the dispersive qubit readout is provided by parametric  {\em radiation} above threshold. This method, illustrated in Fig.~\ref{fig:Rsq_delta_epsilon}, is based on the fact that in the absence of an input signal, $|B|^2=0$, the output signal is zero in the monostable region below the threshold (at blue detunings), $\delta+\delta_q >\sqrt {\epsilon^2 - \Gamma^2}$, while it is finite  above the threshold, 
$|\delta-\delta_q| <\sqrt {\epsilon^2 - \Gamma^2}$,  
where it equals,
$|C|^2 = (2\Gamma/\alpha) \left( -(\delta-\delta_q) + \sqrt{\epsilon^2 - \Gamma^2}\right)$,
according to \Eq{eq:c_hom_stable}. 

The maximum contrast is achieved by choosing the pumping strength, $\epsilon \geq \sqrt{\delta_q^2 + \Gamma^2}$, and the optimum biasing detuning, $\delta \approx  \sqrt {\epsilon^2 - \Gamma^2}- \delta_q$, as illustrated in Fig.~\ref{fig:Rsq_delta_epsilon}. Such a choice guarantees that the blue shifted point, $\delta+\delta_q$, lies in the monostable region close to the threshold, while the red shifted point, $\delta-\delta_q$, lies in the bistable region and not in the tristable region where the trivial cavity state, $|A|=0$, dominates. Then the output radiation power does not depend on $\epsilon$,
\begin{equation}
|C|^2 =  { 4 \Gamma \delta_q\over \alpha}
\,.
\end{equation}
This value is to be compared to the noise value in the monostable region below the threshold. The amplified vacuum noise is given by \Eq{eq:nc_vac_tot} in Sec.~\ref{sec:quantum} and illustrated in the inset of Fig.~\ref{fig:Rsq_delta_epsilon},
\begin{eqnarray}\label{eq:nc_vac_tot_readout}
\ncvac = {\Gamma \epsilon^2 \over  \Gamma^2 + \delta^2 - \epsilon^2 }, \quad B = 0
\,.
\end{eqnarray}
Since the noise diverges at the threshold, the point  $\delta+\delta_q$ is to be chosen not too close to the threshold. It is sufficient to depart from the threshold by $\sim \Gamma$ to have the noise level, $\ncvac \sim \Gamma$. Then for $\delta_q\sim \Gamma$, the radiation to noise contrast becomes,
\begin{equation}
{|C|^2\over \ncvac} \sim {4\Gamma\over \alpha} \gtrsim 100
\,.
\end{equation}

\section{Quantum fluctuations of cavity field and emitted field}
\label{sec:quantum}

So far we discussed the classical regime of parametric resonance in the tunable cavity. In this section, we extend the formalism to the quantum regime, and investigate the quantum properties of the field inside the cavity, and of the output field.  

\subsection{Quantum Langevin equation}
\label{sec:quantum_h}

The Hamiltonian description of the cavity parametric dynamics is a convenient starting point for the extension to the quantum regime. 
To this end we revisit \Eq{eq:Hcav} of Sec.~\ref{subsec:Hamiltonian} and impose canonical commutation relations, 
$[q_n, p_n] = \ui \hbar$, on the conjugated variables of the eigen modes of the closed cavity.
These commutation relations obviously translate to the commutation relations for the resonant variables, $[Q_n, P_n] = \ui \hbar$,
because of the canonical nature of the transformations made in Sec.~\ref{sec:parametric}. The unitary operator, which explicitly defines the corresponding quantum canonical transformation is
\begin{equation}
U(t) = \exp\left(-\iexp \sum_n (q_n^2 + p_n^2) \Omega t/4 \hbar\right)
\,.
\end{equation}
Averaging over rapid oscillations leads to the quantum Hamiltonian coinciding with the one in \Eq{eq:Hcav_RF_QP} with quantum operators replacing respective classical variables.

The quantization of the fundamental mode oscillator implies the quantization of the 
variable $A(t)$ in terms of the conventional commutation relation for the annihilation operator, $[A(t),A^\dag(t)]=1$. 

Due to the linear coupling of the cavity to the transmission line, 
\Eqs{eq:LcavTL_bare}-(\ref{eq:coupl_H}), the input-output formalism outlined in Sec.~\ref{sec:Losses} straightforwardly extends to the quantum regime. To this end, the classical amplitudes of the transmission line modes are to be replaced with the bosonic annihilation and creation operators, with $[A_k(t_0), A_{k'}^\dag(t_0)] = \delta(k-k')$. 
From these commutation relations follows the commutation relation for the incoming field operator,   
$[B(t), B^\dag(t')] = \delta(t-t')$, and similarly for the outgoing field operator $C(t)$.

The scattering relation, \Eq{eq:relation_ABC}, has the same form in the quantum regime, 
\begin{equation}\label{eq:relation_ABC_quant}
C(t) = B(t) - \ui \sqrt{2\Gamma} A(t)
\,,
\end{equation}
while the quantum Langevin equation for the cavity operator $A(t) $ becomes,
\begin{equation}\label{eq:EOM_A_qu}
 \ui \dot A + \delta A + \epsilon A^\dag + \alpha (A^\dag A + 1) A 
+ \ui\Gamma A  =  \sqrt{2\Gamma} B(t)
\,.
\end{equation}
This quantum Langevin equation, together with \Eq{eq:relation_ABC_quant} preserves the
commutation relation $\left[A(t), A^\dag(t)\right] = 1$ for the cavity mode, as shown 
in Appendix~\ref{sec:commutation}.
The conservative part of \Eq{eq:EOM_A_qu} is a dynamical equation associated with the Hamiltonian, 
%
\begin{eqnarray}\label{eq:Hcav_RF}
 &H&
= -\hbar \delta \left(A^\dag A + {1\over2}\right) 
- \frac{\hbar \epsilon}{2} \left(A^{\dag 2} + A^2\right) 
\\
& -& \frac{\hbar \alpha}{2} \left(A^\dag A + {1\over2}\right)^2
+ \hbar \sqrt{2\Gamma} (B A^\dag  + B^\dag A)
\nonumber \,.
\end{eqnarray}
%

\subsection{Small quantum fluctuations}
\label{sec:small_quantum}

The full analytical solution to the nonlinear quantum equation \Eq{eq:EOM_A_qu} is unknown. 
In what follows we restrict to the limit of small quantum fluctuations around the classical stationary states.  Such a restriction is valid far from the bifurcation points and the parametric threshold. Some exact results for the critical fluctuations at such points can be found in literature~\cite{WalMil2008, Drummond2002, KryKhe1996}, also quantum jumps in multistable regimes have been investigated~\cite{DykMalSmeSil1998, Dykman2012}.

To study quantum fluctuations within the framework of a linearized quantum Langevin equation, we assume the cavity field operators to be of the form, $A(t) = A_0 + \hat A(t)$, where $A_0$ is a steady state solution of the classical nonlinear equation, \Eq{eq:EOM_A_cl}, and $\hat A$ describes small quantum fluctuations,
\begin{equation}\label{eq:cond_smallfluct}
 |A_0|^2 \gg \langle \hat A^\dag(t) \hat A(t)\rangle 
\,.
\end{equation}
Similarly, we separate the classical amplitude and quantum fluctuations of the input field in the transmission line, $B(t) = B_0(t) + \hat B(t)$, $\langle \hat B(t) \rangle = 0$.
Then we expand \Eq{eq:EOM_A_qu} around $A_0$ up to linear order in the
quantum fluctuation $\hat A$ to obtain,
\begin{eqnarray}\label{eq:EOM_A_linearized}
\ui  \dot{\hat A} + \znl \hat A + \tilde\epsilon \hat A^\dag
+ \ui\Gamma \hat A
=  \sqrt{2 \Gamma} \hat B 
\,,\\
\label{eq:dnl}
\znl = \zeta + \alpha |A_0|^2 \,, \quad
 \enl = \epsilon + \alpha A_0^2 
\,.\nonumber
\end{eqnarray}
Herein we introduced the effective detuning $\znl$ and the (complex) pump strength $\enl$ by adding the terms proportional to the classical amplitude $A_0$. We note that $A_0$  
 itself depends on the bare parameters $\delta$ and $\epsilon$. 
Quantitatively, the parameter regions where this approximation is valid are identified in Appendix~\ref{sec:validity}.
 
The analysis of \Eq{eq:EOM_A_linearized} goes along the lines of Sec.~\ref{subsec:detuned}, where the response to a classical detuned signal was evaluated.
By introducing Fourier harmonics of the quantum fluctuations in the transmission line,
\begin{equation}
\hat B(\dk) = \int_{-\infty}^{\infty} {d t \over \sqrt{2\pi}} \, \hat B(t) e^{\iexp \dk t},
\end{equation}
and similarly in the cavity, the solution of the linear \Eq{eq:EOM_A_linearized} is cast into the form,
\begin{equation}\label{eq:Aw_Bw}
\left(\begin{array}{l}
  \hat A(\dk)\\ 
  \hat A^\dag(-\dk)
      \end{array}\right)
= \sqrt{2\Gamma}\,
\tilde{\cal A}^{-1} 
\left(\begin{array}{l}
  \hat B(\dk)\\
   \hat B^\dag(-\dk)
      \end{array}\right)
\,,
\end{equation}
where 
\begin{eqnarray}\label{eq:Amatrix}
\tilde{\cal A}^{-1} &=& {1\over \Dnl(\dk)}
\left(\begin{array}{cc}
 \znl - \dk - \ui\Gamma  &  -\enl\\
   -\enl^* & \znl + \dk + \ui\Gamma        
      \end{array}\right), \\
\label{eq:Ddk} 
\Dnl(\dk)\
&=&  (\Gamma -\ui \dk)^2 + \znl^2 - |\enl|^2, \\
|\Dnl(\dk)|^2 &=& \left(\Gamma^2 + \znl^2 - |\enl|^2 - \dk^2 \right)^2 + 4 \Gamma^2 \dk^2
\,, \nonumber
\end{eqnarray}
cf.~\Eq{eq:D(Delta)}.
It follows from this equation, that modes with frequencies $\dk$ and $-\dk$ are coupled 
pairwise by virtue of the parametric pumping.
This property underlines the generation of correlated pairs of photons with frequencies 
$\omega_1+\omega_2= \Omega$, which is analogous to the photon generation under non-degenerate parametric resonance.  

The denominator in \Eq{eq:Aw_Bw} turns to zero at $\dk=0$, if the relation
$\Gamma^2 + \znl^2 - |\enl|^2 = 0$ holds, leading to the divergence of fluctuations at the corresponding parameter values. This happens at the parametric threshold, and at the bifurcation points, and indicates the enhancement of critical fluctuations. 

\subsection{Fluctuations in the cavity}
\label{sec:noiseA}

The full power spectrum of the field in the cavity consists of the sharp line of the amplified (or generated) classical signal, $2\pi |A_0|^2 \delta(\dk)$, together with the noise power spectrum, 
$n_a(\dk)$, 
\begin{eqnarray}
n_a(\dk) 
&=& \int_{-\infty}^{\infty} d \dk' \langle \hat A^\dag(\dk) \hat A(\dk') \rangle  
\,.
\end{eqnarray} 
Solving \Eq{eq:Aw_Bw} and assuming thermal noise in the input field, 
$\left \langle \hat B^\dag(\dk) \hat B(\dk') \right\rangle = N(\dk) \delta(\dk - \dk')$, where 
$N(\dk) = \left(e^{\hbar(\Omega/2 +\dk)/k_B T}  - 1\right)^{-1}$, 
we calculate for the noise power spectrum
\begin{eqnarray}
 n_a(\dk) 
&=& {2\Gamma \over |\Dnl(\dk)|^2 } 
\left\{ 
|\enl|^2 \left[ N\left( - \dk \right) + 1\right] \right.\nonumber \\
&+& \left. ( \Gamma^2 + (\znl - \dk)^2 ) N\left( \dk \right)
\right\}
\,.
\end{eqnarray}

At zero temperature, the noise power spectrum reduces to 
\begin{eqnarray}\label{eq:naw_vac}
 \navac(\dk) = \frac{2\Gamma |\enl|^2}{|\Dnl(\dk)|^2 } 
\,,
\end{eqnarray}
which can be interpreted as the amplified vacuum noise of the input,
manifesting itself as real photons in the cavity.

The noise power spectrum in \Eq{eq:naw_vac} has a resonance structure equivalent
to the resonances in the classical response to a detuned signal discussed in 
Sec.~\ref{subsec:detuned}. The only difference is that now the effective pump parameters enter
\Eqs{eq:Amatrix}--\eqref{eq:Ddk}
instead of the bare pump parameters, 
since we allow here for a finite classical amplitude $A_0$. 
Accordingly, a single resonance at $\dk=0$ is observed under the condition
\begin{equation}\label{eq:cond_oneres}
 \znl^2 < \Gamma^2 + |\enl|^2 
\,,
\end{equation}
and otherwise two resonances are found at
\begin{equation}\label{eq:wres}
\dk = \pm \sqrt{\znl^2 - |\enl|^2- \Gamma^2 }
\,,
\end{equation}
cf.~\Eqs{eq:cond_twores_cl} and \eqref{eq:wres_cl}.

\begin{figure}
\centering
\includegraphics[width=\columnwidth]{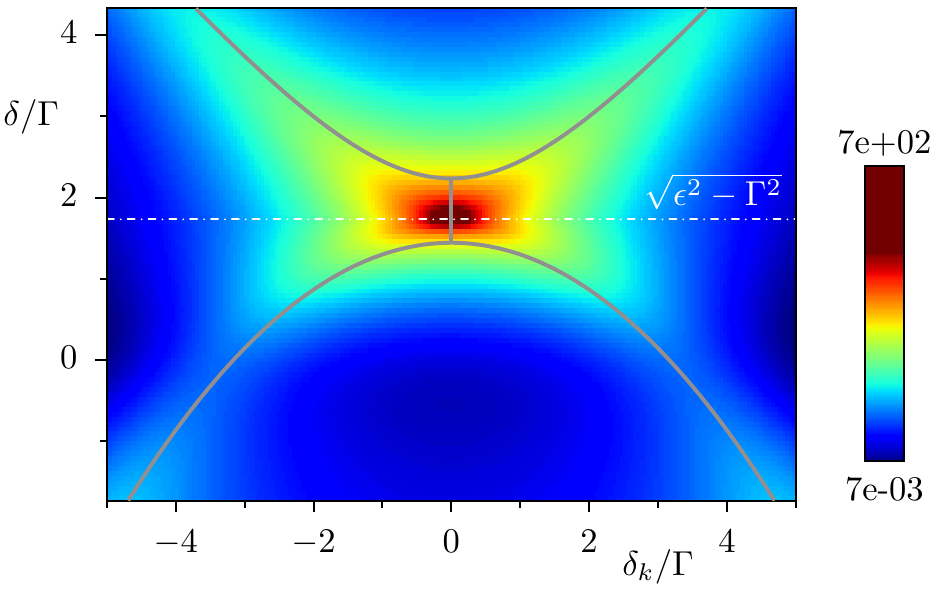}
\caption{
Noise power spectrum $\navac(\dk) \cdot \Gamma$ of the cavity field, Eq.~\eqref{eq:naw_vac},
vs.~pump detuning $\delta$.
For $\delta > \dth$ the classical cavity amplitude is $A_0=0$, 
while $|A_0|^2>0$ for $\delta < \dth$ according to \Eq{eq:a_hom_stable}.
The resonances, Eq.~\eqref{eq:wres}, are indicated by the grey lines.
($\epsilon = 2\Gamma$, $B_0=0$, $\alpha=\Gamma/100$).
}
\label{fig:nafluc_w}
\end{figure}

In Fig.~\ref{fig:nafluc_w} the noise power spectrum $\navac(\dk)$ is presented
as a function of the pump detuning $\delta$ for $\epsilon = 2\Gamma$ and $B_0 = 0$. 
In the monostable regime, $\delta > \dth$, where $A_0 = 0$, the effective pump parameters in  
\Eq{eq:EOM_A_linearized}
are identical to the bare parameters, while in the bistable regime, $\delta < \dth$,
with $A_0$ given by \Eq{eq:a_hom_stable},
they are $\znl = 2\dth - \delta$ and $|\enl|^2 = \delta^2 + \Gamma^2$.
The condition \eqref{eq:cond_oneres} identifies the interval 
$\left(\epsilon^2 -  3\Gamma^2/2 \right)/\dth < \delta < \sqrt{\epsilon^2 + \Gamma^2}$
around the parametric threshold, where the resonance lies at $\dk = 0$.
Outside that interval, once the resonance is split, the separation grows with the 
parameter distance from the threshold,  both below and above the threshold. 
At the parametric threshold itself, $\delta = \dth$, the noise power diverges. 

The total number of photons in the cavity at zero temperature is 
$\langle A^\dag(t) A(t) \rangle = |A_0|^2 + \navac$, with
the noise power
\begin{eqnarray}\label{eq:na_vac}
\navac &=& \int_{-\infty}^{\infty}  { d \dk\over 2\pi} \, \navac(\dk) 
=  {|\enl|^2/2 \over \Gamma^2 + \znl^2 - |\enl|^2}
\,.
\end{eqnarray}
This quantity enters the validity criterium for the linearized Langevin equation, \Eq{eq:cond_smallfluct}, which is analyzed in Appendix~\ref{sec:validity}.

%

\subsection{Fluctuations of the output field}
\label{sec:noiseC}

Similar to the in-cavity field, the full power spectrum of the output field consists of the sharp line of the amplified (generated) classical signal, $2\pi |C_0|^2 \delta(\dk)$, and the noise power spectrum $n_c(\dk)$, 
\begin{eqnarray}\label{eq:ncw_def}
n_c(\dk) 
&=& \int_{-\infty}^{\infty} d \dk' \langle \hat C^\dag(\dk) \hat C(\dk') \rangle  
\,.
\end{eqnarray} 
The relation between the input and output field operators is similar to the one for a detuned classical signal in Sec.~\ref{subsec:detuned}, \Eqs{eq:relation_v_Delta}--\eqref{eq:v_ik},
\begin{eqnarray}\label{eq:relation_v_dk}
\left( \begin{array}{l} \hat C(\dk) \\ \hat C^{\dag}(-\dk)   \end{array} \right)
= 
 \tilde {\cal V}(\dk)
\left( \begin{array}{l} \hat B(\dk) \\ \hat B^{\dag}(-\dk)  
 \end{array} \right) \,, 
\end{eqnarray}
with matrix elements now dependent on the effective pump parameters,
 \begin{eqnarray}\label{eq:tilde_v_ik} 
\tilde v_{11}(\dk) =   (\znl - \ui \Gamma)^2 - \dk^2 - |\enl|^2, \quad
\tilde v_{12} =  2\ui\Gamma\enl
\,, 
\end{eqnarray}
and $\Dnl(\dk)$ is given by \Eq{eq:Ddk}. The matrix elements obey the fundamental relation,
\begin{equation}\label{eq:relation_v11_v12}
|\tilde v_{11}(\dk)|^2 - |\tilde v_{12}(\dk)|^2 = |\Dnl(\dk)|^2
\,,
\end{equation}
which provides the correct commutation relation for the output operators, 
$[\hat C(\dk), \hat C^\dag(\dk')] = \delta(\dk - \dk')$. 

Equation \eqref{eq:relation_v_dk} describes an input-output relation for a linear non-degenerate amplifier \cite{Cav1982} with 
signal and idler modes having frequencies $\dk$ and $-\dk$, respectively, while the input classical tone at $\dk=0$ plays the role of an additional pump. Indeed, the renormalization of the generic pump parameters in 
\Eq{eq:EOM_A_linearized} is an effect of this additional pump
that increases the overall pump strength by $\propto \alpha |A_0|^2$, and also affects the detuning $\delta$ similar to \Eq{eq:omega_shift}. 
We note that \Eq{eq:relation_v_dk} is valid both below and above the threshold, and in the latter case it includes the classical parametric radiation acting as an additional pump signal even in the absence of the classical input.

With the corresponding renormalization of the quantity $\qnl(\dk) = 2\enl \Gamma /\Dnl(\dk)$ that 
characterizes the amplifier gain, \Eq{eq:qw_cl}, we cast the input-output relation, \Eq{eq:relation_v_dk}
into the form,
\begin{eqnarray}\label{eq:Cw_Bw}
&\hat C(\dk)&
 = \sqrt {1+ |\qnl(\dk)|^2} e^{\iexp\tilde \eta(\dk)} \hat B(\dk) + \ui \qnl(\dk) \hat B^\dag(-\dk)
 \nonumber\\
&=& e^{\iexp\tilde \eta(\dk)}\left(\cosh r \hat B(\dk) + \sinh r\, e^{\iexp\chi}\hat B^\dag(-\dk)\right)
,
\end{eqnarray}
where we introduced the standard notation for a non-degenerate parametric amplifier,
\begin{equation}\label{eq:r_squeeze}
\sinh r (\dk) = |\qnl(\dk) |, \; \chi(\dk)  = {\rm arg}\, \qnl(\dk)  -\tilde\eta(\dk)  +{\pi\over 2}
\,.
\end{equation}
The mapping in \Eq{eq:Cw_Bw} is provided by a unitary two-mode squeezing operator~\cite{CavSch1985, KnightBuzek_DruFic2004},
\begin{eqnarray}\label{eq:TMSoperator}
&&\hat C(\dk) =  e^{\iexp\tilde \eta(\dk)}S[\xi] \hat B(\dk) S^\dag[\xi],\\
&&S[\xi] 
=  \exp\left(\int_{0}^\infty \!\! d\dk \left(\xi(\dk) \hat B^\dag(\dk) \hat B^\dag(-\dk) 
- \text{h.c.} \right)\right) 
\,, \nonumber
\end{eqnarray}
where $\xi = re^{i\chi}$. 
This implies  that the stationary state of the output field is a pure state provided the input is a pure state. This is true in spite of  because the evolution of the total system, including the cavity variable, is formally non-unitary due to the presence of the dissipative term in the Langevin equation (\ref{eq:EOM_A_linearized}). 

The noise power spectrum of the output field can be computed from \Eq{eq:Cw_Bw}, and for thermal noise input it reads,  
\begin{eqnarray}
 n_c(\dk)
&=& N( \dk)+ |\qnl(\dk)|^2 
\left[N( \dk) + N( - \dk) + 1\right]
.
\end{eqnarray}
At zero temperature this equation reduces to 
\begin{equation}\label{eq:ncw_vac}
\ncvac(\dk) = |\qnl(\dk)|^2 =2\Gamma \navac(\dk)
\,,
\end{equation}
and describes the generation of real photons from the vacuum under parametric resonance. This phenomenon is closely related to the Dynamical Casimir effect - the creation of real photons from vacuum fluctuations by an accelerated mirror~\cite{Moo1970, Nori2012}.  Here the role of the moving mirror is played by the time-dependent boundary condition, driven by the modulated  magnetic flux through the SQUID. 

The output noise, being proportional to $n_a(\dk)$, inherits the resonant behavior 
of the noise power spectrum in the cavity, as discussed in Sec.~\ref{sec:noiseA} and shown on 
Fig.~\ref{fig:nafluc_w}.  
In the deep subthreshold regime, for very weak pump strength, $\epsilon \ll \Gamma$,
and in absence of an input signal, $B_0 = 0$, \Eq{eq:ncw_vac} takes the form,
\begin{equation}\label{eq:nc_vac_DCE}
 \ncvac(\dk) = {4\epsilon^2\Gamma^2 \over 
\left[\Gamma^2 + (\dk + \delta)^2\right] 
\left[\Gamma^2 + (\dk - \delta)^2\right]}
\,.
\end{equation}
In this limit the resonances move towards $\dk = \pm \delta$, and 
the resonant structure of $\ncvac(\dk)$, approaches the one computed in \cite{JohJohWilNor2010} and observed in~\cite{WilETAL2011}.

\begin{figure}[tb]
\centering
\includegraphics[width=\columnwidth]{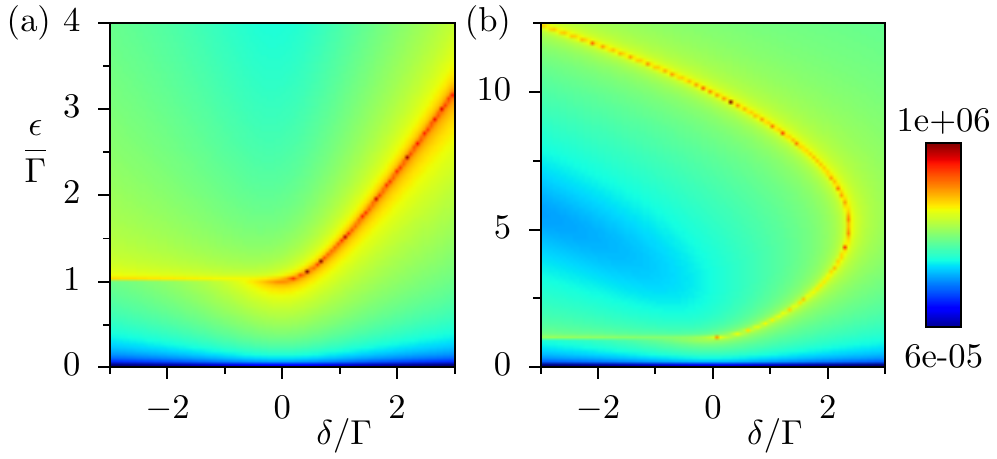}
\caption{
Noise photon flux $\ncvac/\Gamma$, \Eq{eq:nc_vac_tot}, vs.~$\delta$ and $\epsilon$,
assuming (a) bare pump detuning,
and (b) taking into account the pump-induced frequency shift, \Eq{eq:omega_shift}.
($B_0 = 0$, $\alpha=\Gamma/100$).
}
\label{fig:nctot_vac_dpD_eps}
\end{figure}

The  total photon flux in the output field is $\langle C^\dag(t) C(t) \rangle = |C_0|^2 + \ncvac$, 
with the noise photon flux,
\begin{eqnarray}\label{eq:nc_vac_tot}
\ncvac = {\Gamma |\enl|^2 \over  \Gamma^2 + \znl^2 - |\enl|^2 }
\end{eqnarray}
at zero temperature.
Below the parametric threshold, $\epsilon < \sqrt{\delta^2+\Gamma^2}$, the effective parameters in  
\Eq{eq:EOM_A_linearized} are identical to the bare ones, and  \Eq{eq:nc_vac_tot} reduces to 
$\ncvac= \Gamma \epsilon^2 / \left( \Gamma^2 + \delta^2 - \epsilon^2 \right)$.
Above the threshold, $\epsilon > \sqrt{\delta^2+\Gamma^2}$, with $|A_0| > 0$ given by \Eq{eq:a_hom_stable},
\Eq{eq:nc_vac_tot} becomes
%
\begin{equation}
\ncvac = {\Gamma (\delta^2 + \Gamma^2)\over 4\sqrt{\epsilon^2-\Gamma^2} (-\delta + \dth)}
\, .
\end{equation}
The output noise level is illustrated in Fig.~\ref{fig:nctot_vac_dpD_eps} as a function of $\epsilon$ and $\delta$ for $B_0=0$. The right panel demonstrates the effect of back-bending of the threshold line due to the pump-induced frequency shift, 
\Eq{eq:omega_shift} (cf.  Fig.~\ref{fig:homogeneous}(b) in~Sec.~\ref{sec:paramresonance}). The noise is enhanced at the parametric threshold and decreases while moving away from the threshold, there it is estimated as $\ncvac\sim \Gamma$ for $\epsilon,\,\delta \sim \Gamma$. 

Since the noise near the parametric threshold becomes strong, it is useful to evaluate the conditions for the output coherent signal dominating over the noise, 
$ |C_0|^2  \gg \langle \hat C^\dag \hat C \rangle$. 

Above the parametric threshold,  the signal-to-noise ratios are identical for the output field and the field inside the cavity (for $B_0=0$),  
\begin{equation}
{ |C_0|^2  \over \ncvac } = {|A_0|^2 \over \navac  } 
\,.
\end{equation}
Therefore the limitation established by  \Eq{eq:cond_aboveth} for the field in the cavity applies as well to the output field,
\begin{equation}
\epsilon -\Gamma \gg {\Gamma\over8}\left({\alpha\over\Gamma}\right)^{2/3}
\,.
\end{equation}

Below the threshold, the maximum amplified signal is, according to \Eq{eq:c2_b2},
\begin{equation}\label{eq:Csq_quasilin}
 |C_0|^2 \approx  \left( \sqrt{1 + q^2} + q  \right)^2 |B_0|^2
= \frac{(\Gamma + \epsilon)^2}{(\Gamma - \epsilon)^2}|B_0|^2
\,, 
\end{equation}
for $\delta=0$ and $\alpha|A_0|^2\ll \sqrt{\Gamma^2-\epsilon^2}$. Comparing this with \Eq{eq:nc_vac_tot}, we arrive at the constraint on the input signal,
\begin{equation}\label{eq:Bsq_min2}
 |B_0|^2 
\gg {\epsilon^2 \over \Gamma} \frac{1-\epsilon/\Gamma}{(1+\epsilon/\Gamma)^3}
\,.
\end{equation}
This bound is of order $\Gamma$ for $\epsilon\sim \Gamma$, and decreases both at weak pumping and close to the threshold. This is explained, at small $\epsilon\ll \Gamma$, by the fact that the amplification of vacuum noise is small,  while the classical signal remains finite, and, close to the threshold, by the fact that amplification of the signal is more efficient  than the amplification of the noise.  The constraint in \Eq{eq:Bsq_min2} 
is qualitatively similar to the one for the field inside the cavity given by \Eq{eq:Bsq_min1}.

\subsection{Squeezing}
\label{sec:squeezing}

A homodyne detection scheme allows for
measurement of the quadratures of the output signal, and characterization of quadrature fluctuations \cite{WalMil2008, Yurke_DruFic2004, Carmichael2008}. With this method, the output field is mixed with a strong classical field of a local oscillator, $B_{LO}\cos(\Omega t/2 - \theta)$, and the intensity of the mixed signal is measured. This intensity  is proportional to the output field 
quadrature, $I_D(t) = B_{LO} X^\theta(t)$, 
\begin{eqnarray} 
 X^{\theta}(t) &=& C(t) e^{-\ui \theta} + C^\dag(t) e^{\ui \theta}
\,.
\end{eqnarray}
The phase $\theta$ refers to the phase shift of the local oscillator with respect to the parametric pump; variation of $\theta$ allows accessing all the quadratures individually. 

The mean quadrature is determined by the classical output signal 
\begin{equation}\label{eq:X0_theta}
 \langle X^\theta \rangle = X_0^{\theta} 
= C_0 e^{-\ui \theta} + C_0^\ast e^{\ui \theta} 
= 2 |C_0| \cos(\theta_C-\theta)
\,.
\end{equation}
Separating the classical and quantum components, $X^\theta(t) = X_0^\theta + \hat X^\theta(t)$, 
$\hat X^{\theta}(t) =  \hat C(t) e^{-\ui \theta} + \hat C^\dag(t) e^{\ui \theta}$, and 
using the spectral representation of the noise quadratures, 
$\hat X^{\theta}(\dk) = \hat C(\dk) e^{-\iexp \theta} + \hat C^\dag(-\dk) e^{\iexp \theta}$,
we present the corresponding power spectrum in the form 
$2\pi (X_0^\theta)^2 \delta(\dk) + S^\theta(\dk)$, where 
\begin{eqnarray}\label{eq:S_wtheta_def2}
S^{\theta}(\dk) 
&=& \int_{-\infty}^{\infty} d \dk'
\left\langle \hat X^{\theta}(\dk) \hat X^{\theta}(\dk') \right\rangle 
\end{eqnarray}
is the  squeezing power spectrum~\cite{WalMil2008, Carmichael2008}. 
Note that by virtue of the stationary state of the cavity, $\left\langle \hat X^{\theta}(\dk) \hat X^{\theta}(\dk') \right\rangle \propto \delta(\dk + \dk')$, hence only symmetric correlations between the sidebands contribute to the integral, i.e.~the squeezing power characterizes the two-mode squeezing.

We calculate the squeezing power assuming vacuum fluctuations of the input, 
using \Eq{eq:Cw_Bw} for the output field operators.
The result reads,   
\begin{eqnarray}
\!\!\!\! S^{\theta}(\dk) 
&=& 1 + 2|\qnl|^2 - 2\sqrt{1+|\qnl|^2}\, \im\left( \qnl^\ast(-\dk) e^{\iexp (\etanl - 2\theta)} \right) \nonumber\\
\label{eq:S_wtheta}
&=& 1 + {4\Gamma\over|\Dnl(\dk)|^2} \Bigl[
2 \Gamma |\enl|^2 
+ 2 \Gamma \znl \re(\enl e^{-2\ui\theta}) \Bigr. 
\nonumber \\
&& \:\:\: \Bigl. +\:
\im(\enl e^{-2\ui\theta})  \left( \Gamma^2 - \znl^2 + |\enl|^2 + \dk^2 \right)
\Bigr] 
\,.
\end{eqnarray}
Equation $S^{\theta}(\dk) = 1$  corresponds to pure vacuum fluctuations.
The noise squeezing power varies with the phase $\theta$, the maximum and minimum values reached at $\theta_0$ and $\theta_0 + \pi/2$, respectively, with
\begin{eqnarray}\label{eq:theta_max}
\!\!\!\!\!\! \tan( 2\theta_0 ) 
\!=\! \frac{2\znl \Gamma \im(\enl) - [\Gamma^2-\znl^2+|\enl|^2+\dk^2] \re(\enl)}
{2\znl \Gamma \re(\enl) + [\Gamma^2-\znl^2+|\enl|^2+\dk^2] \im(\enl)} 
.
\end{eqnarray}
The corresponding extreme values are determined by the quantity $|\tilde q(\dk)|$, and have the form,
\begin{eqnarray}\label{eq:S_wthetamax}
S^{\theta_0, \theta_0 + \pi/2}(\dk) 
&=& \left( \sqrt{1 + |\qnl(\dk)|^2} \pm |\qnl(\dk)| \right)^2
\,,
\end{eqnarray}
which is similar to the classical gain of the ideal amplifier, \Eq{eq:gain12}, including the relation, 
$S^{\theta_0}(\dk) \cdot S^{\theta_0 + \pi/2}(\dk) = 1$. However, the maximum squeezing and maximum quadrature gain do not generally correspond to the same value of mixing phase $\theta$. Moreover, the amplified classical signal, \Eq{eq:X0_theta}, contains an additional phase, $\theta_C$, which is controlled by the input signal phase $\theta_B$. By varying the latter one may control the signal-to-noise ratio for the quadratures.

\begin{figure}[tb]
\centering
\includegraphics[width=\columnwidth]{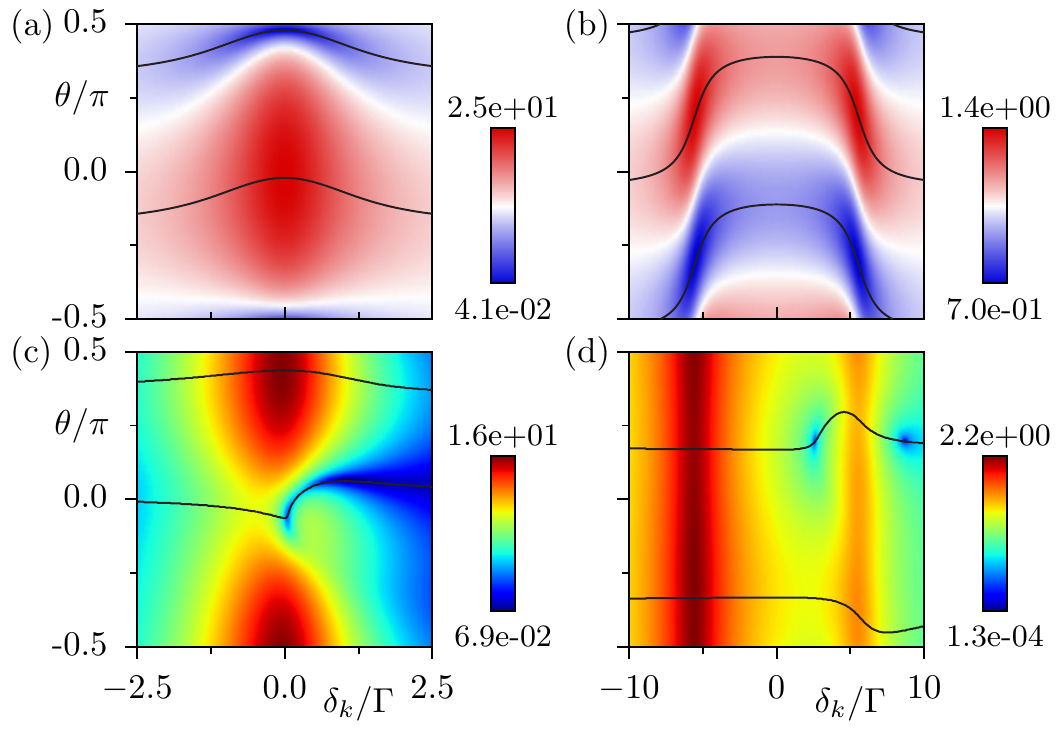}
\caption{
 Squeezing power of output noise 
$S^{\theta}(\dk)$, Eq.~\eqref{eq:S_wtheta} (a-b),
and in-cavity noise $|S_a^{\theta}(\dk)|\cdot \Gamma$, Eq.~\eqref{eq:Sa_wtheta}, (c-d)
for $\epsilon = 1.1 \Gamma$ (left column) and $\epsilon = 3 \Gamma$ (right column).
The quadrature phases for maximum and minimum squeezing power, Eq.~\eqref{eq:theta_max} 
are indicated by the black lines.
($\delta=0$, $B_0=0$, $\alpha=\Gamma/100$).
}
\label{fig:Squeeze_w_theta}
\end{figure}

In Fig.~\ref{fig:Squeeze_w_theta}(a-b) the squeezing power $S^\theta(\dk)$ is shown
for $B_0 = 0$, $\delta=0$, and two different values of $\epsilon > \Gamma$.
The $\theta$-values of maximum and minimum squeezing power are indicated by black lines.

It is useful to also quantify the in-cavity squeezing, by calculating the squeezing power for 
the quadrature operator
$\hat X_A^{\theta}(\dk) = \hat A(\dk) e^{-\ui \theta} + \hat A^\dag(-\dk) e^{\ui \theta}$ in analogy to 
\Eq{eq:S_wtheta_def2}. Although the phase $\theta$ in this case is not related to any externally tunable phase, it might be  relevant for a quadrature-dependent coupling to a qubit placed in the cavity, or to another transmission line.

Assuming vacuum fluctuations in the input field 
we calculate the internal squeezing power using \Eq{eq:Aw_Bw},
\begin{eqnarray}
 S_a^{\theta}(\dk) &=& 
\frac{2\Gamma}{|\Dnl(\dk)|^2}
\Bigl[ \Gamma^2 + (\znl - \dk)^2 + |\enl|^2  - 2 \znl \re(\enl e^{-2 \iexp \theta}) 
\Bigr.
\nonumber \\
\label{eq:Sa_wtheta}
&&\Bigl. - 2 (\Gamma - \ui \dk) \im(\enl e^{-2 \iexp \theta}) \Bigr]
\,.
\end{eqnarray}
Further evaluation of the minimum uncertainty of the cavity quadrature, 
$\langle (\Delta X_a^{\theta})^2 \rangle = (1/2\pi) \int d \dk S_a^{\theta}(\dk)$, results in the value $1/2$, as in the case of linear parametric amplifiers \cite{MilWal1981}, 
i.e.~a factor $1/2$ below the vacuum limit. 

In Figs.~\ref{fig:Squeeze_w_theta}(c-d) $\left|S_a^{\theta}(\dk)\right|$ is illustrated 
for $B_0=0$, $\delta=0$ and $\epsilon>\Gamma$,
in comparison to the external squeezing $S^{\theta}(\dk)$ of Figs.~\ref{fig:Squeeze_w_theta}(a-b).
As a consequence of the effective detuning $\znl$ in Eq.~\eqref{eq:Sa_wtheta}, 
$S_a^{\theta}(\dk)$ is not symmetric around $\dk=0$, as is the case for $S^{\theta}(\dk)$.
%

\subsection{Second order coherence} 
\label{sec:G2}

\begin{figure}[tb]
\centering
\includegraphics[width=\columnwidth]{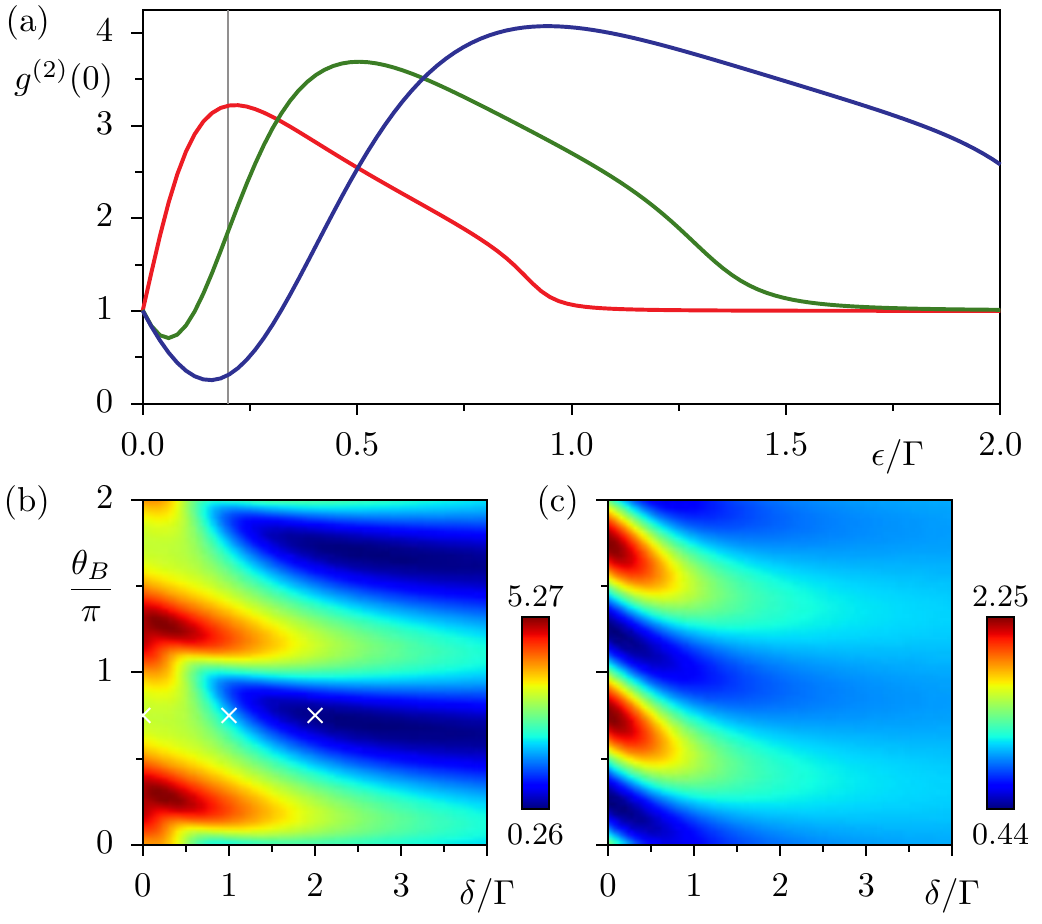}
\caption{
(a) Normalized second order correlation $g^{(2)}(0)$, 
\Eq{eq:G2_t0},
vs.~$\epsilon$ for fixed $\theta_B = -\pi/4$
and for $\delta/\Gamma = 0, 1, 2$ (from bottom to top, the corresponding parameter values are marked with crosses on panel (b)).
(b) $g^{(2)}(0)$ and (c) gain $|C_0|^2/|B_0|^2$ vs.~$\delta$ and $\theta_B$,
for fixed $\epsilon/\Gamma=0.2$ (indicated by the vertical line in (a)).
($|B_0|^2 = \Gamma/10$, $\alpha=\Gamma/100$).
}
\label{fig:g2}
\end{figure}

The two-mode squeezing is a nonclassical property of the amplified noise 
that originates from the production of noise photons in entangled pairs. 
Further information about the nonclassical properties of the correlated output 
photons is provided by a two-photon correlation function, and characteristics of 
two-photon entanglement.

We start with evaluating the second-order correlation function~\cite{WalMil2008},
\begin{eqnarray}
G^{(2)}(\tau) 
&=& \langle C^\dag(t) C^\dag(t+\tau) C(t+\tau) C(t) \rangle 
\,.
\end{eqnarray}
In the presence of the classical output component, this equation takes the form,
\begin{eqnarray}\label{eq:G2_t0_th}
G^{(2)}(\tau) 
&=& |C_0|^4 + 2 |C_0|^2 \Bigl\{ 
\langle \hat C^\dag(t) \hat C(t+\tau) \rangle 
+ \langle \hat C^\dag(t) \hat C(t) \rangle \nonumber\\
&& + \re\left( e^{-2\iexp \theta_C} \langle \hat C(t) \hat C(t+\tau) \rangle \right)
\Bigr\} \nonumber\\  
&& + \langle \hat C^\dag(t) \hat C^\dag(t+\tau) \hat C(t+\tau) \hat C(t) \rangle
\,.
\end{eqnarray}
Explicitly, using \Eq{eq:Cw_Bw}, we obtain for $\tau=0$ and input vacuum noise,
\begin{eqnarray}\label{eq:G2_t0}
 G^{(2)}(0) &=& 
\left(|C_0|^2 + \ncvac\right)^2  \\
&+& \frac{\Gamma}{\Gamma^2 + \znl^2 - |\enl|^2}
 \Bigl[\ncvac (\Gamma^2 + \znl^2 + |\enl|^2)  \nonumber \\
&&  
+ 2 |C_0|^2 \left[ |\enl|^2 + \re\left( \enl (\znl -\ui \Gamma) e^{-2\iexp \theta_C}  \right)\right] \Bigr]
\,. \nonumber
\end{eqnarray}

In Fig.~\ref{fig:g2}(a) the normalized correlation function, 
$g^{(2)}(0) = G^{(2)}(\tau)/\langle C^\dag(t) C(t)\rangle^2 = G^{(2)}(\tau)/(|C_0|^2 + \ncvac)^2$,
is presented as a function of the pumping strength $\epsilon$
for several values of the pump detuning $\delta$.
In the Duffing limit, $\tilde \epsilon = 0$,  all the terms in \Eq{eq:G2_t0} vanish except of the first one,
yielding the coherent state limit, $g^{(2)}(0) = 1$. The same is also true for large pumping strength above the threshold, $\epsilon > \sqrt{\Gamma^2 + \delta^2}$. This is explained by the rapid growth of classical radiation power that dominates over the fluctuations, $|C_0|^2 \gg \ncvac$ (kinks on the curves at $\epsilon/\Gamma>1$). 

At the intermediate pump strengths both bunching ($g^{(2)}(0) > 1$) and antibunching ($g^{(2)}(0) < 1$)
are possible. 
For pure output noise in the absence of classical output, $C_0=0$ (i.e. for $B_0=0$ below the threshold),
only bunching occurs,
$g^{(2)}(0) = 2 + (\Gamma^2 + \znl^2)/|\enl|^2$,
where the degree of bunching exceeds that of classical chaotic radiation, $g^{(2)}(0) > 2$.
This can be interpreted as a consequence of the pair production of noise photons.

When $B_0>0$, also antibunching is possible \cite{ColLou1987} due to the interplay between the classical and the quantum contribution to the correlation, last line in \Eq{eq:G2_t0}. 
It occurs within a relatively narrow window of parameters, $\epsilon < \Gamma$, $\delta \gtrsim \Gamma$,  $|B_0|^2 \approx \Gamma$, 
for which the phase dependence in the last term in \Eq{eq:G2_t0} can introduce a sign change. %

The dependence of $g^{(2)}(0)$ as a function of the input phase $\theta_B$ and the pump detuning $\delta$  is illustrated in Fig.~\ref{fig:g2}(b).
Pronounced antibunching (blue regions) is observed for $|B_0|^2/\Gamma \lesssim 1$, and for those values of $\theta_B$ where the gain approaches unity,
$|C_0|^2/|B_0|^2 \lesssim 1$, compare Fig.~\ref{fig:g2}(c).

\subsection{Two-mode entanglement}
\label{sec:entanglement}

\begin{figure}[tb]
\centering
\includegraphics[width=\columnwidth]{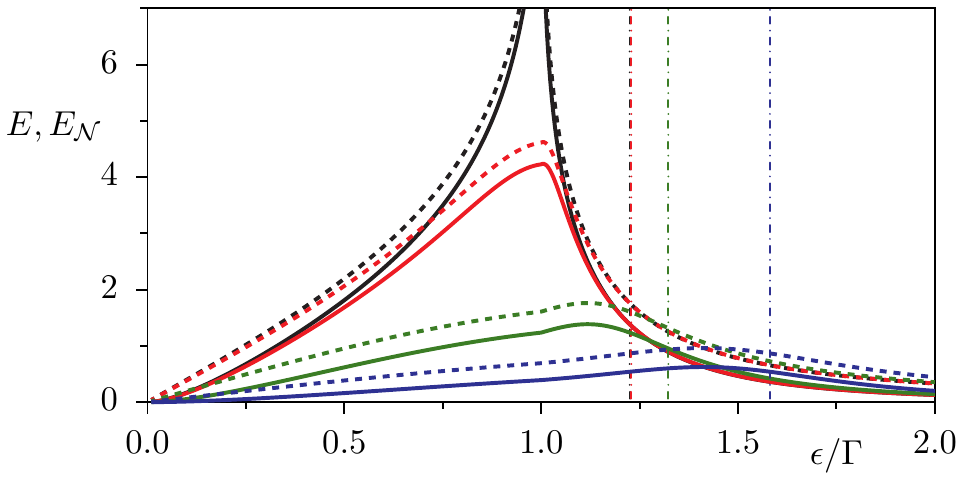}
\caption{
Entanglement entropy $E$ (solid) and logarithmic negativity $E_{\mathcal{N}}$ (dashed) vs.~pump strength $\epsilon$
for $\dk/\Gamma = 0.02, 0.2, 1.0, 2.0$ (from top to bottom).
The vertical lines mark the values of $\epsilon$ at which the resonance, \Eq{eq:wres},
is encountered at the chosen value of $\dk$.
($\delta=0$, $B_0=0$, $\alpha=\Gamma/100$). 
} 
\label{fig:logneg}
\end{figure}

The degree of entanglement between the two modes with frequencies $\dk$ and $-\dk$
can be  quantified with the entanglement entropy \cite{Braunstein2005}, 
\begin{equation}
E(\dk) = -\text{Tr}(\rho(\dk) \ln \rho(\dk)) 
\,,
\end{equation}
where $\rho( \dk)$ is the reduced density matrix of one of the involved modes.
If these modes are entangled, the entropy takes a positive value, $E>0$.

We compute the entanglement entropy for the amplified vacuum noise, using the two-photon wave function of the squeezed state, 
\begin{equation}
|\dk,-\dk \rangle = (\cosh r)^{-1} \sum_{n=0}^{\infty} (\tanh r)^n e^{\iexp n \chi} | n, n \rangle
\,,
\end{equation}
which is obtained by applying the squeezing operator, \Eq{eq:TMSoperator}, to the vacuum input,
$|\dk,-\dk \rangle = S[\xi]\,|0\rangle$, and using the decomposition equation \cite{Collet1988}. 
The reduced density matrix has the form,
 %
\begin{equation}
\rho(\dk) = (\cosh r)^{-2} \sum_{n=0}^{\infty} (\tanh r)^{2 n} | n \rangle \langle n | 
\,,
\end{equation}
giving the entanglement entropy~\cite{vEnk1999}, 
\begin{eqnarray}
 E &=& \cosh^2 r \ln(\cosh^2 r) - \sinh^2 r \ln(\sinh^2 r) 
\,.
\end{eqnarray}
The entropy is nonzero for all $\epsilon > 0$, and follows closely the squeezing parameter $r(\dk)$, asymptotically approaching the linear dependence, $E \approx 2 r$, for $r > 1$.

The entanglement entropy $E$ is shown as function of $\epsilon$ in Fig.~\ref{fig:logneg} (solid lines),
for several values of the detuning $\dk$, and for $B_0 = 0$ and $\delta=0$.
For small detuning, $\dk < \Gamma$, the entropy reaches the maximum at the threshold, $\epsilon=\Gamma$, at which $r(\dk=0)$  [$\qnl(\dk=0)$] diverges. 
With increasing value of the detuning $\dk$ this maximum shifts towards the
value of $\epsilon(\dk)$, at which $\qnl(\dk)$ exhibits the resonance, \Eq{eq:wres}. The entropy rapidly decreases above the threshold, analogous to the behaviour of $g^{(2)}(0)$,  due to the emergence of the classical radiative state, $A_0 \neq 0$, that suppresses $\qnl(\dk)$. 

A convenient measure of entanglement for Gaussian states is provided by the logarithmic negativity \cite{AdeIll2007} related to the covariance matrix for the two entangled modes. The covariance matrix $V_{\alpha \beta}$ is defined through a 4-vector composed of the quadratures,  $R^T = \left(X(-\dk), Y(-\dk), X(\dk), Y(\dk)\right)$,
\begin{equation}
V_{\alpha \beta} = {1\over 2}\langle R_\alpha R_\beta + R_\beta R_\alpha \rangle - \langle R_\alpha \rangle \langle R_\beta \rangle
\,.
\end{equation}
Then splitting the covariance matrix into $2\times2$ submatrices, 
$V = \left(V_1, V_3; V_3^T, V_2\right)$, 
the logarithmic negativity 
is defined as
\begin{equation}
 E_\mathcal{N} = \text{max}\left(0, -\ln(\nu_{-})\right)
\,,
\end{equation}
where $\nu_{-} = \sqrt{ \left(\sigma - \sqrt{\sigma^2 - 4 \det V} \right)/2 }$
and $\sigma = \det V_1 + \det V_2 - 2 \det V_3$.
For entangled states the logarithmic negativity takes positive values. 

For amplified vacuum noise we obtain a simple result, using \Eq{eq:Cw_Bw}, 
\begin{eqnarray}\label{eq:logneg_}
 E_\mathcal{N} 
=  2\ln \left(\sqrt{1+|\qnl(\dk)|^2} + |\qnl(\dk)| \right) = 2r
\,,
\end{eqnarray}
i.e.~the logarithmic negativity is equal to twice the squeezing parameter $r(\dk)$.

The logarithmic negativity $E_\mathcal{N}$ is shown in Fig.~\ref{fig:logneg} with dashed lines.
Its functional behavior is basically equivalent to that of the entropy $E$.  

Our calculation shows that the degree of the two-mode entanglement is significantly enhanced 
in the presence of the parametric resonance. To evaluate the exact maximum entanglement value one needs to go beyond the quasilinear approximation and include the nonlinear effect. We make a qualitative estimate  by taking the function $|\qnl(\dk)|$ at the threshold, $\epsilon=\Gamma$, $\delta=0$, 
and at $\dk=0$, and for  the cavity field given by \Eqs{Amax} and \eqref{Amin} 
assuming input power, $|B_0|^2 \sim \Gamma$, corresponding to one photon per bandwidth. This yields an  estimate, 
\begin{equation}
E_{\rm max}\approx E_{{\cal N}{\rm max}} \sim {\rm const}\cdot \ln{\Gamma\over\alpha}
\,,
\end{equation}
with a numerical constant of order one. This crude estimate seems to  agree  with more accurate evaluation of the critical fluctuations \cite{KryKhe1996}. For values $\Gamma/\alpha \sim 100$ achievable in tunable cavities, the entanglement entropy can accordingly reach the values $4.5 - 5$. 
This is significantly larger than the values calculated \cite{JohETAL_logneg2012} for a non-resonant open transmission line with modulated boundary, and also exceeds the  values reported for experimental parametric Josephson devices \cite{EichlerPRL2011,FluETAL2012}.

\section{Summary}
\label{sec:summary}

We have developed a consistent theory of parametric resonance in a high quality tunable superconducting cavity. 
We considered the  nonlinear classical dynamics of the cavity  both below and above the parametric threshold, and analyzed amplification of external signals, and parametric radiation. We also studied quantum properties of the amplified and radiative fields.

The non-equidistance of the cavity frequency spectrum enabled us to formulate the theory of the degenerate parametric resonance in terms of the one  encountered in a nonlinear parametric oscillator. We identified the parameters of this effective oscillator as functions of the cavity  generic characteristics, and investigated the multistable cavity dynamics in a relevant range of the effective parameters.

The operation of the device in the monostable regime as a nonlinear parametric amplifier is characterized with a phase-dependent differential gain, which increases at small input power and reaches the maximum value at the parametric threshold. We found that this maximum value scales with the ratio of the damping coefficient and the nonlinearity coefficient, $\Gamma/\alpha$. We also found that the relation between the maximum and minimum gain for an ideal linear amplifier is violated in the nonlinear regime, $G_{{\rm min}}\ll 1/G_{{\rm max}}$. Extremely small values of 
$\alpha$ available in tunable cavities allows for very large gain and strong amplification vs. deamplification contrast. 

Amplification of detuned signals was found to exhibit sideband resonances within a specific region of the cavity parameters. This effect can be used for enhancing the amplification bandwidth while maintaining high gain.

The application of the device as a parametric bifurcation amplifier was discussed in regard to dispersive qubit readout. The advantage of the parametric regime compared to the conventional JBA  is a high sensitivity of  the strength of the output signal to the variation of the cavity frequency. This, together with a high amplification gain, provides a potential for improving the fidelity of qubit single shot readout. 

Yet another suggested method for qubit readout is based on a high contrast between the strengths of parametric radiation above the threshold and amplified noise below the threshold.

Small-amplitude quantum fluctuations around the classical signal were investigated for the in-cavity field and the output field. The limit of small fluctuations is appropriate in a wide range of the device parameters except of small regions of critically enhanced fluctuations close to the bifurcation points and the parametric threshold. The theory is analogous to the one for a quantum linear amplifier. The strength of the amplified noise increases in the vicinity of the threshold in accord with the classical gain. The same is also true for the two-mode squeezing and the entanglement quantified with the entanglement entropy and the logarithmic negativity. 
At the threshold, the estimated magnitude of the squeezing parameter may reach the values of a few units, exceeding that achievable e.g.~in non-resonant Josephson mixers. 

The second order coherence is dominated by strong bunching for small classical inputs, resulting from the production of noise photons in pairs. However, for classical inputs with strength comparable to the vacuum noise, significant antibunching is predicted resulting from the interference of the classical and quantum field components.

To conclude, we note that the developed theory straightforwardly extends to the regime of non-degenerate parametric resonance, when the pumping frequency is commensurate with a combination of cavity resonances. 
Similarly, in this case, strongly enhanced amplification gain is to occur near the parametric threshold, as well as strongly enhanced two-mode squeezing and entanglement of the cavity modes selected by the resonance. 

Yet another extension of the theory is readily done for a two-sided cavity parametrically pumped by two SQUIDs, attached to both sides of the cavity  \cite{IdaMaria_thesis}. The dynamics of this device is equivalent to the  single-sided parametric cavity, provided the SQUIDs are operated at the same pump frequency. The parametric resonance is then controlled by an effective pump strength, which depends on the phase shift between the actual pumps. For equal pump amplitudes the parametric effect is maximum for the out-of-phase pumping (``breathing'' mode), while for the in-phase  pumping (``translational'' mode) the parametric instability is completely suppressed.\\

{\it Acknowledgement} 
We acknowledge useful discussions with Chris Wilson, Per Delsing, G\"oran Johansson, Konrad Lehnert, and Tim Duty.
Support from FP-7 IP SOLID is gratefully acknowledged.


\begin{appendix}{}

\section{Lagrangian and boundary condition of the flux-tunable cavity}
\label{sec:SQUID}

In this appendix we derive the Lagrangian of the flux-tunable cavity, Eq.~\eqref
{eq:Lcav_x_cos_orig0}, and give arguments for its validity. 

We start with a description of the SQUID establishing the connection between the cavity and the pump line. The generalized coordinates of the SQUID are the superconducting phase $\phi_d = \phi(x=d)$ at the cavity edge $x=d$, 
the phase $2\!f$ dropping over the inductance $L$ of the SQUID loop, and the phase $2\!\fext$ dropping over the coupling inductance $L_{\text{ext}}$ of the pump line, see Fig.~\ref{fig:device}.

The SQUID is modelled as symmetric, with two identical Josephson junctions, each having a Josephson energy $E_J$ and a capacitance $C_J$. To simplify notation we assume that the SQUID is grounded in such a way that its geometric inductance $L$ is divided into two equal parts $L/2$, with a phase drop of $f$ over each part. Thus, the phase difference on one of the Josephson junctions is $\phi_d-f$ and $\phi_d+f$ on the other. The coupling to the flux line is inductive, with a mutual inductance $M \ll L, L_{ext}$. The full SQUID Lagrangian is
\begin{eqnarray}
\mathcal{L}_{S} 
&=& 
\phantom{-}\fq \left( 
  \frac{C_J}{2} \left(\dot \phi_d - \dot f\right)^2 
+ \frac{C_J}{2} \left(\dot \phi_d + \dot f\right)^2 
\right) \nonumber\\
&&+ E_{J} \cos\left(\phi_d - f\right)  
+ E_{J} \cos\left(\phi_d + f\right) \nonumber\\
&& - \fq \frac{1}{2 L} \left( 4 f^2 + 8 {M \over L_{ext}} f \fext \right)
,
\end{eqnarray}
or, written with the capacitive energy of a Josephson junction $E_C = (2e)^2/(2 C_J)$
and the inductive energy of the SQUID loop $E_L = (\hbar/2e)^2/(2 L)$ 
\begin{eqnarray}\label{eq:LSQUID}
\mathcal{L}_S
&=& \phantom{-} \frac{\hbar^2}{2 E_C} \dot \phi_d^2 + 2 E_{J} \cos\left(\phi_d\right) \cos{f} 
\nonumber\\
&&
+ \frac{\hbar^2}{2 E_C} \dot f^2  - E_L \left( 4 f^2  + 8 \frac{M}{L_{ext}} f \fext \right)
.
\end{eqnarray}
Separating the $\phi_d$-dependent terms (first line) from the purely $f$-dependent ones (second line), 
$\mathcal{L}_S = \mathcal{L}_{S}[\phi_d, f] + \mathcal{L}_{S}[f]$, 
the former can be combined with the bare cavity Lagrangian
\begin{equation}\label{eq:Lcav0}
 \mathcal{L}^{(0)}_{\text{cav}} 
= \frac{d \ELcav}{2 v^2} \int_0^d d x \left(\dot \phi^2 - v^2 \phi'^2 \right)
,
\end{equation}
with the inductive energy of the cavity $\ELcav = \hbar^2/( L_0 d (2e)^2)$.
Together, these form the Lagrangian $\mathcal{L}_{\text{cav}}$
of the flux-tunable cavity, Eq.~\eqref{eq:Lcav_x_cos_orig0}.

For typical cavity and junction dimensions the orders of the three inductive energies in the Lagrangian,
Eqs.~\eqref{eq:LSQUID}-\eqref{eq:Lcav0}, are distinctly different.
The dominant energy, $E_L/ \hbar \sim 10^{5} \uGHz$, determined by the small geometric inductance of the SQUID loop ($L \approx 10^{-12} \uH$), is larger than  the Josephson energy of the SQUID, $2 E_J/\hbar \sim 10^{4} \uGHz$, and that dominates over the inductive energy of the cavity, 
$\ELcav/\hbar \sim 400 \uGHz$ (for $d L_0 \approx 2 \cdot 10^{-9} \uH$). 
Furthermore, the Josephson plasma frequency $\omega_J = \sqrt{2 E_J E_C}/\hbar \sim 300 \uGHz$ is high compared to the fundamental cavity resonance, $\omega_0 \sim 40 \uGHz$
(compare Sec.~\ref{subsec:CavModes}).

The equations of motion for $\phi_d$ and $f$, 
according to the full Lagrangian $\mathcal{L}_{\text{cav}} + \mathcal{L}_{S}[f]$,
\begin{eqnarray}
\label{eq:EOM_phid_orig}
&&\!\!\!
\frac{\hbar^2}{E_C} \ddot \phi_d + 2 E_J \cos{\!f} \sin\phi_d + \ELcav d \phi'_d  = 0\\
\label{eq:EOM_phiL_orig}
&&\!\!\!
\frac{\hbar^2}{2 E_C} \ddot f +\! E_J \cos\phi_d \sin{\!f} 
+\! 4 E_L \left(\! f + \frac{M \fext}{L_{ext}} \!\right) 
\!= 0
\end{eqnarray}
describe two coupled nonlinear oscillators. For $\fext=0$ the equilibrium is $(f=0, \phi_d=0)$.

In general, the coupled dynamics of nonlinear, driven oscillators features chaotic behaviour. We restrict our analysis to the case $\phi_d \ll 1$ and assume that this is fulfilled even in the presence of a resonant excitation by the external field $\fext(t)$. Under this condition the equation of motion for $f$,  \Eq{eq:EOM_phiL_orig}, decouples from the other oscillator,
\begin{eqnarray}\label{eq:EOM_phiL_2} 
\frac{\hbar^2}{2 E_C} \ddot f  + E_J \sin{f} + 4 E_L \left( f + \frac{M}{L_{ext}} \fext \right) &=& 0
\,,
\end{eqnarray}
and the dynamical equation for $\phi_d$, \Eq{eq:EOM_phid_orig}, then depends only parametrically
on $f(t)$, cf.~\Eq{eq:EOM_phid}.

We suppose the external force of the form
$\fext(t) = F_{\text{ext}} + \df_{\text{ext}}(t)$, $\df_{\text{ext}} \ll 1$,
and separate the SQUID phase response $f(t) = F + \df(t)$ 
into a constant equilibrium shift $F$, governed by the equation,
\begin{eqnarray}\label{eq:phiL_eq}
E_J \sin{F} + 4 E_L \left( F + (M/L_{ext}) F_{ext} \right) = 0
\,,
\end{eqnarray}
and a small harmonic oscillation, $\df(t) \ll 1$, driven by $\df_{\text{ext}}(t)$,
\begin{eqnarray}\label{eq:varphiL}
\!\!\!\!\!\!\frac{\hbar^2}{2 E_C} \delta\!{\ddot f} + (E_J \cos{F} + 4 E_L) \df 
=- \frac{4 E_LM}{L_{ext}} \df_{\text{ext}}(t) 
.
\end{eqnarray}
Assuming
$\df_{\text{ext}}(t) = \df_{\text{ext}} \cos{\Omega t}$, we write the stationary solution in the form, 
$\df(t) = \df \cos{\Omega t}$,
\begin{equation}
\df = -{8 E_L M E_C/\hbar^2 L_{ext} \over \omega_f^2 - \Omega^2} \, \df_{\text{ext}}
\,,
\end{equation}
where $\omega_f = \omega_J \sqrt{\cos{F} + 4 E_L/E_J}$
is the frequency of the $f$-oscillator, which is much larger than the frequency of the pump, 
$\Omega \approx 2\omega_0 $.

Linearized around the equilibrium shift $F$, 
Eq.~\eqref{eq:EOM_phid_orig} becomes
\begin{eqnarray}\label{eq:boundaryR1_nlin_}
\frac{\hbar^2}{E_C} \ddot \phi_d &+& 2 E_J \left[ \cos{F} - \sin{F} \df \cos(\Omega t) \right] \sin\phi_d
\nonumber\\ 
&+& \ELcav d \phi'_d  = 0
\,.
\end{eqnarray}
For $\df = 0$ this boundary condition determines the cavity mode spectrum, 
Eqs.~\eqref{eq:phi_n} and~\eqref{eq:dispersion}.

Further expanding \Eq{eq:EOM_phid_orig} to the second order with respect to $\df$ 
leads to the pump induced shift of the cavity frequencies. Indeed, averaging over time, we get a correction to the Josephson energy, $2E_J\cos F (1 - \df^2/4)$, which will modify  
\Eq{eq:dispersion} accordingly. In particular, for the fundamental mode we get from \Eq{eq:w0_approx},
\begin{eqnarray}\label{eq:pump_frshift}
 {\omega_0\,(\df) \over \omega_0\,(0)} &\approx& 
1 - \frac{\gamma \df^2}{4}
\,.
\end{eqnarray}
One could also expand \Eq{eq:EOM_phiL_orig} to the second order with respect to $\df$, which would lead, after the time averaging, to a shift of the static bias $F$, and eventually to an additional shift of the cavity frequencies. However, this effect is small, by virtue of the parameter $E_J/E_L\ll1$, compared to the shift (\ref{eq:pump_frshift}).

\section{Mode representation of cavity Lagrangian}
\label{sec:Lcav_n}

In this appendix, we express the Lagrangian of the flux-tunable cavity, \Eq{eq:Lcav_x_cos_orig0}, 
in the mode representation, \Eq{eq:Lcav_n_nlin},
based on the expansion~\eqref{eq:modeexpansion} of the cavity field. 

Firstly, making use of \Eq{eq:dispersion}, the overlap integrals of the non-orthogonal modes are
\begin{eqnarray}
\label{eq:nonorth1}
 \lefteqn{\int_0^d d x \cos{k_n x} \cos{k_m x} = } \qquad\\
&& {d M_n \over 2}  \delta_{nm} - {2 C_J \over C_0} \cos{k_n d} \cos{k_m d} 
\nonumber\\
\label{eq:nonorth2}
\lefteqn{ \int_0^d d x  k_n k_m \sin{k_n x} \sin{k_m x} 
={d k_n^2 M_n \over 2} \delta_{nm}} \qquad\\
&&- {2 C_J k_m^2 \over C_0} \cos{k_n d} \cos{k_m d} - k_m \cos{k_n d} \sin{k_m d} 
\nonumber
\,,
\end{eqnarray}
where we have defined the coefficients $M_n$, \Eq{eq:Mn}.
With these, the bulk contribution to the cavity Lagrangian 
becomes
\begin{eqnarray}
\label{eq:Lbulk_n}
\lefteqn{
\fq {C_0 \over 2} 
\int_0^d d x \left(\dot \phi^2 - v^2 (\phi')^2 \right)
=} \\
&&\frac{1}{2} \sum_n \left[ M_n \dot q_n^2 - M_n v^2 k_n^2 q_n^2 \right] \nonumber\\
&+& \frac{1}{2} \sum_{n,m} \Bigl[
- {2C_J \over C_0} \cos{k_n d} \cos{k_m d} \dot q_n \dot q_m   \Bigr.\nonumber\\
&&\Bigl. + v^2  \cos{k_n d} \left({2C_J \over C_0} k_m^2 \cos{k_m d} + k_m \sin{k_m d} \right) q_n q_m
\Bigr]
\nonumber \,.
\end{eqnarray}

In the remaining boundary contribution of \Eq{eq:Lcav_x_cos_orig0}, we firstly separate a time dependent, nonlinear potential term 
\begin{eqnarray}
 V(\phi_d, t) = -2 E_J \left( \cos{f(t)} \cos{\phi_d} + \cos{F} {\phi_d^2 \over 2} \right)
\end{eqnarray}
from the harmonic contribution,
\begin{eqnarray}
\lefteqn{ 
\fq {2 C_J \over 2} \dot \phi_d^2 + 2 E_J \cos{f(t)} \cos{\phi_d} = } \qquad \\ 
&& \fq {2 C_J \over 2} \dot \phi_d^2 - 2 E_J \cos{F} {\phi_d^2 \over 2} - V(\phi_d, t)
\,. \nonumber
\end{eqnarray}
The mode-representation, \Eq{eq:modeexpansion}, of the harmonic part becomes, using
$(2e)^2/(\hbar^2 C_0) = v^2/(d \ELcav)$,
\begin{eqnarray}\label{eq:Lbound_harmonic}
\lefteqn{\fq {2 C_J \over 2} \dot \phi_d^2 - 2 E_J \cos{F} {\phi_d^2 \over 2} =} \qquad \\
&& {2 C_J \over 2 C_0} \sum_{n,m} \cos{k_n d} \cos{k_m d} \,\dot q_n \dot q_m \nonumber \\
&-& {2 E_J v^2 \cos{F} \over 2 \ELcav d} \sum_{n,m} \cos{k_n d} \cos{k_m d} \,q_n q_m 
\,. \nonumber
\end{eqnarray}
The first term of this cancels directly with a term in the bulk contribution, \Eq{eq:Lbulk_n}.
Further, using the definition of the modes in \Eq{eq:dispersion}, we note that
\begin{equation}
k_m \left({2C_J \over C_0} \!\cos{k_m d} + \sin{k_m d}\right) \!=\! {2 E_J\! \cos{F} \over \ELcav d}\! \cos{k_m d}
,
\end{equation}
leading to further cancellation of terms between the bulk and the boundary contribution.
The remaining terms are
\begin{eqnarray}
\mathcal{L}_{\text{cav}} = \frac{1}{2} \sum_n \left[ M_n \dot q_n^2 - M_n v^2 k_n^2 q_n^2 \right]  - V(q_n,t)
\,,
\end{eqnarray}
with $V(q_n,t) = V(\phi_d(q_n), t)$. 
This is the mode representation of the cavity Lagrangian in \Eq{eq:Lcav_n_nlin}.

\section{Transmission line amplitudes and scattering relation}\label{sec:Langevin}

In this appendix we show the relation of the flux amplitude $B$ introduced in \Eq{eq:Bt}
to the incoming field, and similarly for the flux amplitude $C$ of the outgoing field,
as well as their mutual relation given in \Eq{eq:relation_ABC}.
The incoming and outgoing fields in the transmission line are defined, respectively, as
\begin{eqnarray}
\label{eq:phiin}
\!\!\!\!\!\! \phi_{in}(t)  \!&=&\!
{e\over \hbar} \sqrt{ \!\frac{\hbar}{\pi C_0}} 
\!\int_0^{\infty} \!\!\!\! {d k \over \sqrt{\omega_k}} 
\!\!\left[ a_k(t_0) e^{-\iexp \omega_k (t-t_0)} \!+\! \text{h.c.}
\right] \\
\label{eq:phiout}
\!\!\!\!\!\! \phi_{out}(t) \!&=&\!
{e\over \hbar} \sqrt{ \!\frac{\hbar}{\pi C_0}} 
\!\int_0^{\infty} \!\!\!\! {d k \over \sqrt{\omega_k}} 
\!\!\left[ a_k(t_1) e^{-\iexp \omega_k (t-t_1)} \!+\! \text{h.c.}
\right] 
.
\end{eqnarray}
These are based on the solutions of \Eq{eq:EOM_ak}, which is expressed in terms of initial amplitudes $a_k(t_0)$ 
at a time $t_0 < t$ in the past, or in terms of final amplitudes $a_k(t_1)$ 
at a time $t_1 > t$ in the future,
\begin{eqnarray}
\label{eq:a_kt0}
\lefteqn{ a_k(t) = a_k(t_0) e^{-\iexp \omega_k (t-t_0)}  } \qquad \\
&&+ {C_c \over C_0 d} \sqrt{2 d \omega_0 \omega_k \over M_0 \pi \hbar} 
\int_{t_0}^t d t' p(t') e^{-\iexp \omega_k (t-t')} \nonumber \\
\label{eq:a_kt1}
\lefteqn{ a_k(t) = a_k(t_1) e^{-\iexp \omega_k (t-t_1)} } \qquad \\
&&- {C_c \over C_0 d} \sqrt{2 d \omega_0 \omega_k \over M_0 \pi \hbar} 
\int_{t}^{t_1} d t' p(t') e^{-\iexp \omega_k (t-t')}  \nonumber 
.
\end{eqnarray}
Such a definition is justified, as will be shown below, by attributing different propagation directions along the transmission line for the incoming and outgoing field components, which can be separated by circulators and hence have physical meaning.

We firstly use the solution \eqref{eq:a_kt0} to evaluate the field in the transmission line, Eq.~\eqref{eq:phiTL} with $q_k = \sqrt{\hbar/2} (a_k + a_k^\dag)$,
\begin{eqnarray}
\lefteqn{ \phi_{TL}(x,t) \!=\! 2e\!\!
\int_0^{\infty} \!\! \frac{d k \cos{k x}}{\sqrt{\pi \hbar C_0 \omega_k}}
\left[ a_k(t_0) e^{-\iexp \omega_k (t-t_0)}  \!+\! \text{h.c.} \right] } \\
&& + {4e C_c \over \pi \hbar C_0 d}\sqrt{2 d \omega_0 \over C_0 M_0}
\int_{t_0}^t \!\!d t' p(t') \int_0^\infty \!\!\!d k \cos(\omega_k (t-t')) \cos{kx}
\nonumber .
\end{eqnarray}
The contribution from the first line can be straightforwardly identified with $\phi_{in}$ from Eq.~\eqref{eq:phiin}, and equals
$\phi_{in}(t-x/v) + \phi_{in}(t+x/v)$.
The integral in the second line is evaluated (for $x<0$)
\begin{eqnarray}
\lefteqn{
 \int_0^{\infty} d k 2 \cos(\omega_k(t-t')) \cos(k x) } \nonumber  \\
&=& \int_0^{\infty} d k \left[ \cos(k v(t-t') - kx) + \cos(k v(t-t') + kx) \right] \nonumber \\
&=& \frac{d}{d t} \int_0^\infty \frac{d k}{kv} 
\left[ \sin(k v(t-t') - kx) + \sin(k v(t-t') + kx) \right] \nonumber\\
&=& \frac{\pi}{2v} \frac{d}{d t} 
\left[ \text{sgn}(t-t' - x/v) + \text{sgn}(t-t' + x/v) \right] \nonumber \\
&=& \frac{\pi}{v}
\left[ \delta(t-t' - x/v) + \delta(t-t' + x/v) \right] 
.
\end{eqnarray}
The second $\delta$-function gives a contribution at $t'=t+x/v$, 
whereas the first, for $x<0$, is not included in the integration limits, $t' = t-x/v > t$,
and therefore
\begin{equation}
\int_0^{\infty} \!\! d k \!\int_{t_0}^t \!d t' p(t') \cos(\omega_k(t-t')) \cos(k x) 
 = \frac{\pi}{2v} p(t+x/v)
.
\end{equation}
Taken together, the field in the transmission line reads
\begin{eqnarray}\label{eq:phiTM_1}
\phi_{TL}(x,t) &=&  \phi_{in}(t-x/v) + \phi_{in}(t+x/v)  \nonumber \\
&+& {2e \over \hbar} {C_c \over C_0 d}  \sqrt{2 d \omega_0 \over C_0 M_0} {1 \over v} p(t+x/v)
\,.
\end{eqnarray}
Alternatively, 
the transmission line field, \Eq{eq:phiTL}, 
can be evaluated from the second solution for the $a_k(t)$, \Eq{eq:a_kt1}, yielding 
\begin{eqnarray}\label{eq:phiTM_2}
 \phi_{TL}(x,t) &=&  \phi_{out}(t-x/v) + \phi_{out}(t+x/v)  \nonumber \\
&-& {2e \over \hbar} {C_c \over C_0 d}  \sqrt{2 d \omega_0 \over C_0 M_0} {1 \over v} p(t-x/v)
\,.
\end{eqnarray}
By subtracting Eqs.~\eqref{eq:phiTM_1} and \eqref{eq:phiTM_2} at $x=0$
we can establish a relation between the incoming and the outgoing field components,
\begin{eqnarray}\label{eq:phi_inout0}
 \phi_{out}(t) - \phi_{in}(t)
=
{2e\over\hbar} {C_c \over C_0 d} \sqrt{2 d \omega_0 \over M_0 C_0} { p(t) \over v}
.
\end{eqnarray}
%
Note that the last term can also be expressed by the derivate of the cavity field at $x=0$, 
$\dot \phi_0(t) \approx (2e/\hbar) \sqrt{2\omega_0/(C_0 d M_0)} p(t)$, 
using $p(t) \approx \omega_0 \dot q(t)$ in the weak coupling approximation.

We can now evaluate \Eq{eq:phi_inout0} at $t+x/v$, and insert in Eq.~\eqref{eq:phiTM_1}, 
such that the transmission line field is expressed as a linear combination of $\phi_{in}$ and $\phi_{out}$ alone,
\begin{equation}
  \phi(x,t) =  \phi_{in}(t-x/v) + \phi_{out}(t+x/v) 
,
\end{equation}
demonstrating the role of $\phi_{in}$ and $\phi_{out}$ as incoming and outgoing field components.

Finally, we want to relate the general input-output relation, \Eq{eq:phi_inout0},
with the slow varying amplitudes of the resonant approximation.
To that end we separate the fast time oscillation with frequency $\Omega/2$ in \Eqs{eq:phiin} and (\ref{eq:phiout}),  
\begin{eqnarray}
\!\! \phi_{in}(t) &=& {2e \over \hbar} \sqrt{\hbar \over 2 C_0 \omega_0 v}
\left( B(t) e^{-\iexp \Omega t/2} + \text{h.c.} \right) \\
\!\! \phi_{out}(t) &=& {2e \over \hbar} \sqrt{\hbar \over 2 C_0 \omega_0 v}
\left( C(t) e^{-\iexp \Omega t/2} + \text{h.c.} \right) 
,
\end{eqnarray}
where 
\begin{eqnarray}\label{eq:def_B}
\!\!\!\!    B(t) &=& \sqrt{\omega_0 v \over 2\pi} e^{\iexp \Omega t/2}
\int_0^{\infty} \!\!{d k \over \sqrt{\omega_k}}
a_k(t_0) e^{-\iexp \omega_k (t-t_0)} \\
\!\!\!\!  C(t) &=& \sqrt{\omega_0 v \over 2\pi} e^{\iexp \Omega t/2}
\int_0^{\infty} \!\!{d k \over \sqrt{\omega_k}}
a_k(t_1) e^{-\iexp \omega_k (t-t_1)} 
\,.
\end{eqnarray}
Within the resonant approximation, $\sqrt{\omega_k}\approx \sqrt{\omega_0}$, $a_k e^{i\Omega t/2} = A_k$, these quantities coincide with the ones defined in Sec.~\ref{sec:Losses}, cf.~\Eq{eq:Bt}.
The cavity momentum is expressed in the rotating frame as well, 
\begin{eqnarray}\label{eq:p_t}
 p(t) 
&=& -\ui \sqrt{\hbar \over 2} \left( A(t) e^{-\iexp \Omega t/2} - A^\dag(t) e^{\iexp \Omega t/2} \right)
\,,
\end{eqnarray}
with the slowly time-dependent cavity amplitude $A(t)$.
By setting these expressions into  \Eq{eq:phi_inout0},
multiplying with  $e^{\iexp \Omega t/2}$ and averaging over fast oscillation, the corresponding input-output relation is obtained in the rotating frame, cf.~\Eq{eq:relation_ABC}.
\begin{equation}
  C(t) - B(t) =  {-\ui C_c \over C_0 d} {\sqrt{d \omega_0 \Omega} \over \sqrt{2 M_0 v}} A(t)
= -\ui \sqrt{2\Gamma_0} A(t)
\,.
\end{equation}
%

\section{Commutation relations}
\label{sec:commutation}

In this appendix we show that the quantum Langevin equation, \Eq{eq:EOM_A_qu},
preserves the commutation relation of the cavity amplitude.
To this end we express the solution of \Eq{eq:EOM_A_qu}
in terms of the propagator $U(t,t_0) = \exp(-\ui H_{\text{cav}} (t-t_0)/\hbar)$ 
with the Hamiltonian that governs the dynamics of the isolated cavity, 
\begin{eqnarray}
 H_{\text{cav}}
&=& -\hbar \delta \left(A^\dag A + {1\over2}\right) 
- \frac{\hbar \epsilon}{2} \left((A^\dag)^2 + A^2\right) \\
&&-\frac{\hbar \alpha}{2} \left(A^\dag A + {1\over2}\right)^2 
\nonumber
\end{eqnarray}
(cf.~\Eq{eq:Hcav_RF}).
The operator $A$ refers to the Schr\"odinger picture
and coincides initially with the Heisenberg operator $A(t_0)=A$.
At time $t$ the solution is
\begin{eqnarray}
A(t) &=& e^{-\Gamma (t-t_0)} U^{-1}(t,t_0) A(t_0) U(t,t_0) \\
&-& \ui \sqrt{2\Gamma} 
\int_{t_0}^{t} d t' e^{-\Gamma (t-t')} U^{-1}(t,t') B(t') U(t,t') \nonumber \\
&=& e^{-\Gamma (t -t_0)} U^{-1}(t,t_0) A(t_0) U(t,t_0)  \\
&-& \ui \sqrt{2\Gamma} \int_{t_0}^{t} d t' e^{-\Gamma (t-t')} B(t') 
\,, \nonumber 
\end{eqnarray}
where we have used the fact that $H_{\text{cav}}$ and $B(t')$ commute since $A$ is  uncorrelated with the operators $a_k(t_0)$ of the incoming transmission line modes of which $B(t)$ is composed. 
Using this solution we are able to evaluate the equal time commutator,
\begin{eqnarray}
&& \left[ A(t), A^\dag(t) \right]
= e^{-2\Gamma (t-t_0)} \bigl[\, U^{-1}(t,t_0) A(t_0) U(t,t_0)\,,\bigr. \\
&& \hspace*{2cm}
\bigl. U^{-1}(t,t_0) A^\dag(t_0) U(t,t_0) \,\bigr] +\ui \sqrt{2\Gamma} e^{-\Gamma (t-t_0)}\nonumber\\
&& \hspace*{0.5cm}\times \int_{t_0}^t \!\!d t' e^{-\Gamma t'} 
\Bigl(
\left[U^{-1}(t,t_0) A(t_0) U(t,t_0),\,  B^\dag(t') \right] \Bigr. \nonumber\\
&& \phantom{\hspace*{0.5cm}\times \!\int_{t_0}^t \!\!d t' e^{-\Gamma t'} }
- \left[B(t'),\, U^{-1}(t,t_0) A^\dag(t_0) U(t,t_0) \right] \Bigr) \nonumber\\
&&\hspace*{0.5cm}+ 2\Gamma \iint_{t_0}^t d t' d t'' e^{-\Gamma(t'+t'')} 
\left[ B(t'),\, B^\dag(t'') \right]  
\,.
\end{eqnarray}

The mixed commutators vanish, again with the argument of initially uncorrelated cavity and transmission line operators, leaving
\begin{eqnarray}
\lefteqn{\left[ A(t), A^\dag(t) \right] =} \nonumber\\
&&\hspace*{0.5cm} e^{-2\Gamma (t-t_0)} U^{-1}(t,t_0) \left[  A(t_0),\, A^\dag(t_0) \right] U(t,t_0) \nonumber\\
&&\hspace*{0.5cm} + 2\Gamma \iint_{t_0}^t d t' d t'' e^{-\Gamma(t'+t'')}  
\left[ B(t'),\, B^\dag(t'') \right] 
\,.
\end{eqnarray}
Finally, using $\left[  A(t_0),\, A^\dag(t_0) \right] = 1$
and $\left[ B(t'),\, B^\dag(t'') \right] = \delta(t'-t'')$,
we arrive at the desired result,
\begin{equation}
\left[ A(t), A^\dag(t) \right] = e^{-2\Gamma (t-t_0)}
- \left(e^{-2\Gamma (t-t_0)} - 1 \right) = 1
\,.
\end{equation}
The invariance of the commutation relation under the Langevin evolution, \Eq{eq:EOM_A_qu},
follows from the correct combination of the damping term, $\Gamma A$,
and the fluctuations in the amplitude $B$. Averaging over fluctuations would violate the exact unitary evolution and break the commutation relation.

\section{Validity of quantum linearized treatment}
\label{sec:validity}

Having evaluated the magnitude of the quantum fluctuations, we are able to discuss the region of validity of the linearized equation, \Eq{eq:EOM_A_linearized}. Two assumptions have been made for the derivation: the amplified signal at frequency $\delta_k=0$ has been treated as a classical field, $|A_0|^2 \gg 1$, and its magnitude to exceed the amplified external noise, $|A_0|^2 \gg n_a$. Together these conditions are (cf. \Eq{eq:cond_smallfluct}),
\begin{equation}\label{eq:cond_smallfluct2}
|A_0|^2 \gg \text{max}\left(1,\;n_a\right)
\,.
\end{equation}
We analyze these conditions separately above and below the threshold, 
at zero temperature, and at $\delta=0$ for simplicity.

Above the threshold, $\epsilon>\Gamma$, the parametric radiation dominates over the input signal. Neglecting the input, $B_0=0$, we have $|A_0|^2 = \sqrt{\epsilon^2-\Gamma^2}/\alpha$ in accord with \Eq{eq:a_hom_stable}.  Then $|\enl|^2 = \Gamma^2$ and $\znl = 2\dth$, and the amplified vacuum noise is,
\begin{equation}
 \navac = {\Gamma^2/8 \over \epsilon^2 - \Gamma^2} 
\,.
\end{equation}
The conditions of \Eq{eq:cond_smallfluct2},
\begin{equation}\label{eq:cond_aboveth}
 {\alpha \over \Gamma} \ll 
 \left\{ \begin{array}{ll}
\displaystyle \left({\epsilon^2 \over \Gamma^2} - 1\right)^{1/2},
& \displaystyle{\epsilon\over\Gamma} > {3\over 2\sqrt{2}} \\
\displaystyle  8 \left({\epsilon^2 \over \Gamma^2} - 1\right)^{3/2}, 
& \displaystyle 1 < {\epsilon\over \Gamma} < {3\over 2\sqrt{2}} 
        \end{array}\right.
\end{equation}
are fulfilled everywhere except of the close vicinity of the threshold. 
Near the threshold the  external noise dominates, while its role diminishes with growing pump strength. In the limit of very strong pumping, $\epsilon \gg \Gamma$, \Eq{eq:cond_aboveth} reduces to $\alpha/\epsilon \ll 1$. This result can be understood from a purely Hamiltonian argument. The semiclassical limit requires the quantum uncertainty of a state localized in a quantum well,  $\sim\hbar$, to be  much smaller than the total phase-space volume of the well. The latter can be estimated from the separatrix area, $\propto \hbar\epsilon/\alpha \gg \hbar$, 
cf.~second inset of Fig.~\ref{fig:homogeneous}(a).
Since in the semiclassical limit tunneling between the wells is exponentially suppressed,
it is consistent to treat noise as local fluctuations in each well separately.

Below the threshold, $\epsilon<\Gamma$, \Eq{eq:cond_smallfluct2} imposes constraints on the input field $B_0$. To be consistent with the linear description of fluctuations, we consider the quasilinear limit of the classical  response, 
$\alpha |A_0|^2 \ll\sqrt{\Gamma^2-\epsilon^2}$. With this assumption, the maximum magnitude of the field in the cavity,  \Eq{eq:a_nonlin} with $\theta_B=-\pi/4$, 
reads
\begin{eqnarray}
 |A_0|^2 
\approx {2 \Gamma_0  \over (\Gamma - \epsilon)^2} |B_0|^2
\,,
\end{eqnarray}
while the amplified vacuum noise is
\begin{equation}
 \navac \approx {\epsilon^2/2 \over \Gamma^2 - \epsilon^2}
\,.
\end{equation}
The number of amplified vacuum photons inside the cavity is small at weak pumping but grows and passes the one-photon level at $\epsilon = \sqrt{2/3}\,\Gamma$, and becomes dominant while approaching the parametric threshold. Using these estimates we extract from \Eq{eq:cond_smallfluct2} the lower bound on the input signal,
\begin{equation}\label{eq:Bsq_min1}
 {|B_0|^2 \over \Gamma} \gg 
\displaystyle \left\{ \begin{array}{ll}
\displaystyle \frac{ (\Gamma - \epsilon)^2}{2\Gamma^2}\,,
&\displaystyle  {\epsilon\over \Gamma} < \sqrt{2\over3} \\
\displaystyle {\epsilon^2 \over 4\Gamma^2} \frac{\Gamma - \epsilon}{\Gamma + \epsilon}\,,
& \displaystyle  \sqrt{2\over3} < {\epsilon\over \Gamma} < 1
        \end{array}\right.
\,.
\end{equation}
For very small pump strength, $\epsilon\ll \Gamma$, the constraint (\ref{eq:Bsq_min1}) reduces to $|B_0|^2 \gg \Gamma$, 
which is qualitatively similar to  a high quality Duffing cavity, 
in which the resonant field fed by the input $|B_0|^2 \gg \Gamma$ achieves a large (classical) value, 
$|A_0|^2 \gg 1$. The quasilinear approximation in this case, $\alpha \ll \alpha |A_0|^2 \ll \Gamma$ is valid as soon as $\alpha \ll \Gamma$, and it imposes an upper bound on the input,
$|B_0|^2 \ll \Gamma^2/2\alpha$. 

Close to the threshold, the amplified signal grows with $\epsilon$  more rapidly than the noise, and remains dominant at practically all input signals. This regime persists until the nonlinear effect breaks the quasilinear approximation at $\alpha |A_0|^2 \sim \sqrt{\Gamma^2-\epsilon^2}$, and the signal amplitude saturates. The corresponding constraint on the input reads, 
\begin{equation}\label{eq:lim_quasilin}
{|B_0|^2\over \Gamma} \ll  {\Gamma\over \sqrt{2} \alpha}
\left(1-{\epsilon\over \Gamma}\right)^{5/2}, \quad \Gamma-\epsilon \ll \Gamma
\,.
\end{equation}
In terms of $\epsilon$, the upper bound for this regime is given by the condition
\begin{equation}
1-{\epsilon\over \Gamma} \ll \left({\sqrt 2\over 8}{ \alpha \over \Gamma} \right)^{2/3}
\,.
\end{equation}
For experimentally relevant cavity parameters, $\alpha/\Gamma < 1/10$, 
our estimates for the relative noise strength
are therefore valid up to $\epsilon \sim 0.95\Gamma$.

\end{appendix}


\end{document}